\documentclass[11pt]{article}
\usepackage{teschool09,epsfig}

\def\Journal#1#2#3#4{{#1} {\bf #2}, #3 (#4)}

\def\NIMA{{\em Nucl. Instrum. Methods} A}

\def\PLB{{\em Phys. Lett.}  B}

\def\be{\begin{equation}}
\def\ee{\end{equation}}
\def\bea{\begin{eqnarray}}
\def\eea{\end{eqnarray}}

\begin{document}
\centerline{\underline {{\bf \Large arXiv: 1008.3736 [physics.ins-det]} 26 April 2010}}
\vspace*{1cm}
\title{Gaseous Detectors: recent developments and applications\footnote{Invited Lectures 
at the Trans-European School of High Energy 
Physics, Zakopane, Poland, July 8-14 (2009); 
http://events.lal.in2p3.fr/TESchool09/}}

\author{\Large\bf Maxim Titov}
\address{CEA Saclay, DSM/IRFU/SPP, 91191 Gif sur Yvette, France}

\maketitle\abstracts{
 Since long time, the compelling scientific goals of future high energy physics experiments 
were a driving factor in the development of advanced detector technologies.
 A true innovation in detector instrumentation concepts came in 1968, with the development 
of a fully parallel readout for a large array of sensing elements - the Multiwire 
Proportional Chamber (MWPC), which earned Georges Charpak a Nobel prize in physics in 1992.
Since that time radiation detection and imaging with fast gaseous detectors, capable of economically 
covering large detection volume with low mass budget, have been playing an important role 
in many fields of physics.
Advances in photo-lithography and micro-processing techniques in the chip industry during 
the past decade triggered a major transition in the field of gas detectors from wire structures 
to Micro-Pattern Gas Detector (MPGD) concepts, revolutionizing cell size limitations 
for many gas detector applications.
The high radiation resistance and excellent spatial and time resolution make them an 
invaluable tool to confront future detector challenges at the next generation of colliders.
The design of the new micro-pattern devices appears suitable for industrial production. 
 Novel structures where MPGDs are directly coupled to the
CMOS pixel readout represent an exciting field
allowing timing and charge measurements as well as precise
spatial information in 3D.
Originally developed for the high energy physics,
MPGD applications has expanded to nuclear physics,
UV and visible photon detection, astroparticle and neutrino physics, 
neutron detection and medical physics.
}

\section {Gaseous Detectors: Historical Overview}

 The single wire proportional counter, invented more 100 years ago by
E. Rutherford and H. Geiger~\cite{geiger_1908}, and its high gain successor,
the Geiger-Mueller counter first described in 1928~\cite{geiger_1928},
can be considered the ancestors of all modern gaseous detectors and were
for many decades a major tool for the study of ionizing radiation.
Forty years ago, in 1968, the instrumentation 
in experimental particle physics was revolutionized 
with the advent of the Multi-Wire Proportional Chamber (MWPC)~\cite{charpak_1968}.
 With its excellent accuracy and modest rate capability, the MWPC allowed large detector 
areas to be instrumented with fast tracking detectors and were able
to localize particle trajectories with sub-$mm$ precision.
Confronted by the increasing demands of particle physics experiments, MWPCs
have continuously improved over the 
years~\cite{cern_7709,annrev_1984_34_285,cern_2004_040,Grupen_1996,Blum_2008}.
 Gradually replacing slower detectors, numerous generations of gaseous devices,
with novel geometries and exploiting various gas properties,
has been developed: Drift Chamber~\cite{walenta_1971}, 
Multi-Drift Module~\cite{Bouclier88}, JET Chamber~\cite{Drumm80},
Time Projection Chamber (TPC)~\cite{Nygren78}, Time Expansion Chamber~\cite{tec_1979},
Multi-Step Chamber~\cite{multistep_1978}, Ring-Imaging Cherenkov Counter~\cite{rich_1977}
and many others.
However, limitations have been reached in maximum rate capability and detector granularity.
A fundamental rate limitation of wire chambers, due to positive ion accumulations,
was overcome with the invention of Micro-Strip Gas Chamber (MSGC)~\cite{msgc_1988}, 
capable of achieving position
resolution of few tens of microns at particle fluxes exceeding $MHz/mm^2$~\cite{barr_1998}.
Developed for the projects at high-luminosity colliders,
MSGC's filled a gap between the performing but expensive solid-strip 
detectors and cheap but rate-limited traditional wire chambers.
Despite their impressive performance, detailed studies of long-term
behavior at high rates have revealed two possible weaknesses of the MSGC 
technology: a slow degradation under sustained irradiation (aging),
and appearance of rare but destructive discharges in presence of
highly ionizing particles~\cite{bagaturia_2002}.

The invention of Micro-Pattern Gas Detectors (MPGD), in particular the Gas Electron 
Multiplier (GEM)~\cite{GEM_NIMA386}, the Micro-Mesh Gaseous Structure 
(Micromegas)~\cite{Micromegas_NIMA376}, and
other micro pattern detector schemes, offers the potential to develop new gaseous 
detectors with unprecedented spatial resolution, high rate capability, large sensitive 
area and operational stability~\cite{annrev_1999_49_341,nima581_25}. 
Recent developments in radiation hardness research with state-of-the-art MPGD's revealed
that they might be even less vulnerable to radiation-induced performance
degradation than standard silicon micro-strip detectors~\cite{titov_ieee_tns49_1609,arXiv_0403055}.
In some applications, requiring 
very large-area coverage with moderate spatial resolutions, more coarse macro-patterned 
detectors, e.g. thick-GEMs (THGEM)~\cite{thgem_nima478_377,thgem_nima535_303,thgem_nima598_107} 
or patterned resistive thick GEM devices (RETGEM)~\cite{retgem_nima581_225} could offer 
an interesting and economic solution. 
In addition, the availability of highly 
integrated amplification and readout electronics allows for the design of gas-detector 
systems with channel densities comparable to that of modern silicon 
detectors. 
 An elegant solution is the use of a CMOS pixel 
ASIC\footnote{Application Specific Integrated Circuit}, assembled directly below the 
GEM or Micromegas amplification 
structures~\cite{nature411_2001,bellazzini_NIMA535_477,campbell_NIMA540_295,bamberger_NIMA573_361}. 
Modern wafer post-processing also allows for the integration of Micromegas grid
directly on top of a CMOS pixelized readout chip~\cite{chefdeville_NIMA556_490}. 
 In 2008, the RD51 collaboration at CERN has been established to further advance technological
developments of micro-pattern gaseous detectors and associated electronic-readout systems,
for applications in basic and applied research~\cite{rd51webpage}.

\section{Basic Principles: Ionization, Transport Phenomena and Avalanche Multiplication}

 The process of detection in gas proportional counters starts with 
the inelastic collisions between the incident particles 
and gas molecules.
These collisions lead to excitation of the medium (followed 
by the emission of the light, the basis of scintillation
detectors) and ionization, the primary signal for tracking devices.
 The number $N_P$ and the space distribution of the primary ionization
clusters depend on the nature and energy of the radiation. 
 The primary electrons can often have enough energy to further ionize the
medium; the total number of 
electron-ion pairs ($N_T$) is usually a few times larger than the
number of primaries ($N_P$).
 Table~\ref{gasproperties} provides values of relevant parameters in some commonly
used gases at NTP (normal temperature and pressure) for unit charge
minimum-ionizing particles (MIPs)\cite{PDG_2008}.
Values often differ, depending on the source, and those in the table should be taken only
as approximate.

\begin{table}[htb]
\caption{ Properties of noble and molecular gases at normal temperature and pressure 
(NTP: 20$^\circ$ C, one atm). $E_X$, $E_I$: first excitation, 
ionization energy; $W_I$: average energy per ion pair; $dE/dx|_{\rm min}$, 
$N_P$, $N_T$: differential energy loss, primary and total number of 
electron-ion pairs per $cm$, for unit charge minimum ionizing particles.}
\vspace{0.4cm}
\begin{center}
\begin{tabular}{|c|c|c|c|c|c|c|c|}
\hline
Gas & Density & $E_x$ & $E_I$ &  $W_I$ & $dE/dx|_{\rm min}$ & $N_P$ & $N_T$ \\
  & mg cm$^{-3}$ & eV &  eV   &  eV& keV cm$^{-1}$ & cm$^{-1}$ & cm$^{-1}$ \\
\hline
  He           & 0.179 & 19.8 & 24.6 & 41.3 & 0.32 & 3.5 & 8 \\
\hline
  Ne           & 0.839 & 16.7 & 21.6 & 37 & 1.45 & 13 &  40 \\
\hline
  Ar           & 1.66  & 11.6 & 15.7 & 26 & 2.53 & 25 &  97 \\
\hline
  Xe           & 5.495 &  8.4 & 12.1 & 22 & 6.87 & 41 & 312 \\
\hline
  CH$_4$       & 0.667 &  8.8 & 12.6 & 30 & 1.61 & 28 &  54 \\
\hline
  C$_2$H$_6$     & 1.26  &  8.2 & 11.5 & 26 & 2.91 & 48 & 112 \\
\hline
  iC$_4$H$_{10}$ & 2.49  &  6.5 & 10.6 & 26 & 5.67 & 90 & 220 \\
\hline
  CO$_2$       & 1.84  &  7.0 & 13.8 & 34 & 3.35 & 35 & 100 \\
\hline
  CF$_4$       & 3.78  & 10.0 & 16.0 & 54 & 6.38 & 63 & 120 \\
\hline
\hline
\end{tabular}
\end{center}
\label{gasproperties}
\end{table}

 The primary statistics determines several intrinsic performances
of gas detectors, such as efficiency, time resolution and 
localization accuracy.
 The actual number of primary interactions follows the Poisson's statistics; 
the inefficiency of a perfect detector with a thin layer of gas 
is given by exp(-$N_P$).
 For example, in one $mm$ of $Ar/CO_2$ (70:30) approximately 6~$\%$ of all MIPs
do not release a single primary electron cluster and therefore can not be detected.
The total energy loss, sum of primary and secondary ionization, follows 
a statistical distribution described by a Landau function, with characteristic
tails towards higher values.
A simple composition law can be used for gas mixtures: for example, the
number of primary ($N_P$) and total ($N_T$) electron-ion pairs produced 
by MIP in a 1 cm of $Ar/CO_2$ (70:30) mixture at NTP:
\begin{equation}
N_P = 25 \cdot 0.7 + 35 \cdot 0.3 = 28~\frac{pairs}{cm};~~~
N_T = \frac{2530}{26}\cdot 0.7 + \frac{3350}{35}\cdot 0.3 \approx 97~\frac{pairs}{cm} \\
\label{number_np_nt}
\end{equation}

 While charged particles release ionization trail of primary 
electron clusters, low energy $X$-rays undergo a single localized interaction, usually 
followed by the emission of the photo-electron, accompanied by the 
lower-energy photon or Auger electron.
 For example, a 5.9~keV $X$-ray converts in argon mainly on a $K$ shell
($3.2~keV$); the emitted photo-electron with energy $E_{\gamma}-E_K \sim 2.7~keV$
has a practical range in detector of $\sim 200~\mu m$. 
In addition, with 85~$\%$ probability another (Auger) electron with energy 
$\sim 3.2~keV$ ($\sim 250~\mu m$ range in argon) is ejected; in the remaining 
cases, a $3~keV$ K-L fluorescence photon is produced with a mean absorption
length of 40~$mm$.
 The sum of the energies of photo-electron and Auger electron is responsible
for the main 5.9~keV peak, while fluorescence mechanism leads to the 
$Ar$ escape peak.
 The total number of electron-ion pairs created by $X$-ray
absorbed in argon can be evaluated by dividing its energy by
the $W_I$: $\frac{5900}{26} \approx 227$.

 Once released in the gas, and under the influence of an applied electric
field, electrons and ions drift in opposite directions and diffuse towards
the electrodes.
 When electron move through the gas, the magnitude of the scattering cross
section in an electron-atom (molecule) collision
is determined by the details of atomic and molecular structure.
Therefore, the drift velocity and diffusion of electrons depend very
strongly on the nature of the gas, namely on the structure of
 inelastic cross-section, arising from
rotational and vibrational levels of molecules.
In noble gases, the inelastic cross section is zero below excitation and ionization
thresholds.
Large drift velocities are achieved by adding polyatomic gases
(usually CH$_4$, CO$_2$ or $CF_4$), having large inelastic cross sections at
moderate energies, which results in ``cooling" electrons into an energy
range of the Ramsauer-Townsend minimum (located at $\sim$0.5~eV) of the
momentum transfer elastic cross-section of argon. The reduction in both the total
electron scattering cross-section and the electron energy results in a
large increase of electron drift velocity (for a compilation of
electron-molecule cross sections see Ref.~\cite{XSECTIONS}). Another
principal role of the polyatomic gas is to absorb the ultraviolet (UV)
photons emitted by the excited inert gas atoms. The quenching of UV
photons occurs through the photo-decomposition of polyatomic molecules.
Extensive collections of experimental data\cite{Peisert84} and theoretical
calculations based on transport theory\cite{biagi_nima421_234} permit estimates
of drift and diffusion properties in pure gases and their mixtures.
In a simple approximation, gas kinetic theory provides the following relation
between drift velocity, $v$, and the mean collision time between electron
and molecules,
$\tau$ (Townsend's expression): $v=eE\tau/m$.
Values of drift velocity and diffusion for some commonly used
gases at NTP are given in Fig.~\ref{transverse_parameters}a and
Fig.~\ref{transverse_parameters}b.
These have been computed with the MAGBOLTZ program\cite{SIMULATION}.
Using fast $CF_4$-based mixtures at fields around $kV/cm^{-1}$, 
the electron drift velocity is around 10~$cm \cdot \mu s^{-1}$ and
the corresponding drift time 100~$ns \cdot cm^{-1}$.
For different conditions, the horizontal axis must be scaled inversely with the gas
pressure or density, $1/P$, where $P$ is the pressure. Standard
deviations for longitudinal ($\sigma_L$) and transverse diffusion
($\sigma_T$) are given for one cm of drift, and scale with the square root
of distance.
Since the collection time is inversely proportional to the drift
velocity, diffusion is smaller in gases having large drift velocities,
such as for CF$_4$.
In the presence of an external magnetic field, the Lorentz
force acting on electrons between collisions deflects the drifting electrons
and modifies the drift properties. 
For parallel electric
and magnetic fields, drift velocity and longitudinal diffusion are not
affected, while the transverse diffusion can be strongly reduced:
$\sigma_T (B) = {\sigma_T (B=0)}/{\sqrt{1+\omega^2\tau^2}}$.
The dotted line in Fig.~\ref{transverse_parameters}b represents $\sigma_T$
for the classic $Ar/CH_4$ (90:10) mixture at 4~T.
This reduction is exploited in TPC to improve spatial resolution.

\setlength{\unitlength}{1mm}
\begin{figure}[bth]
 \begin{picture}(55,55)
 \put(-5.0,-5.0){\includegraphics{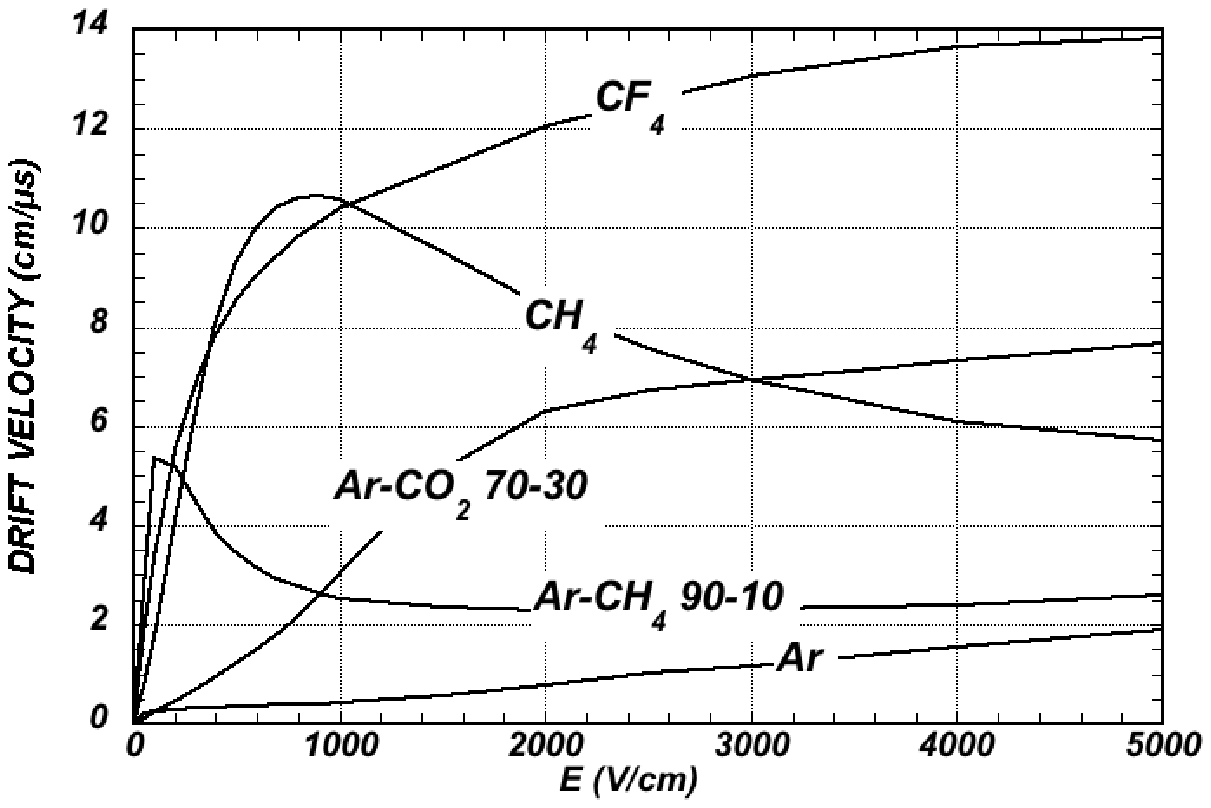}}
 \put(73.0,-5.0){\includegraphics{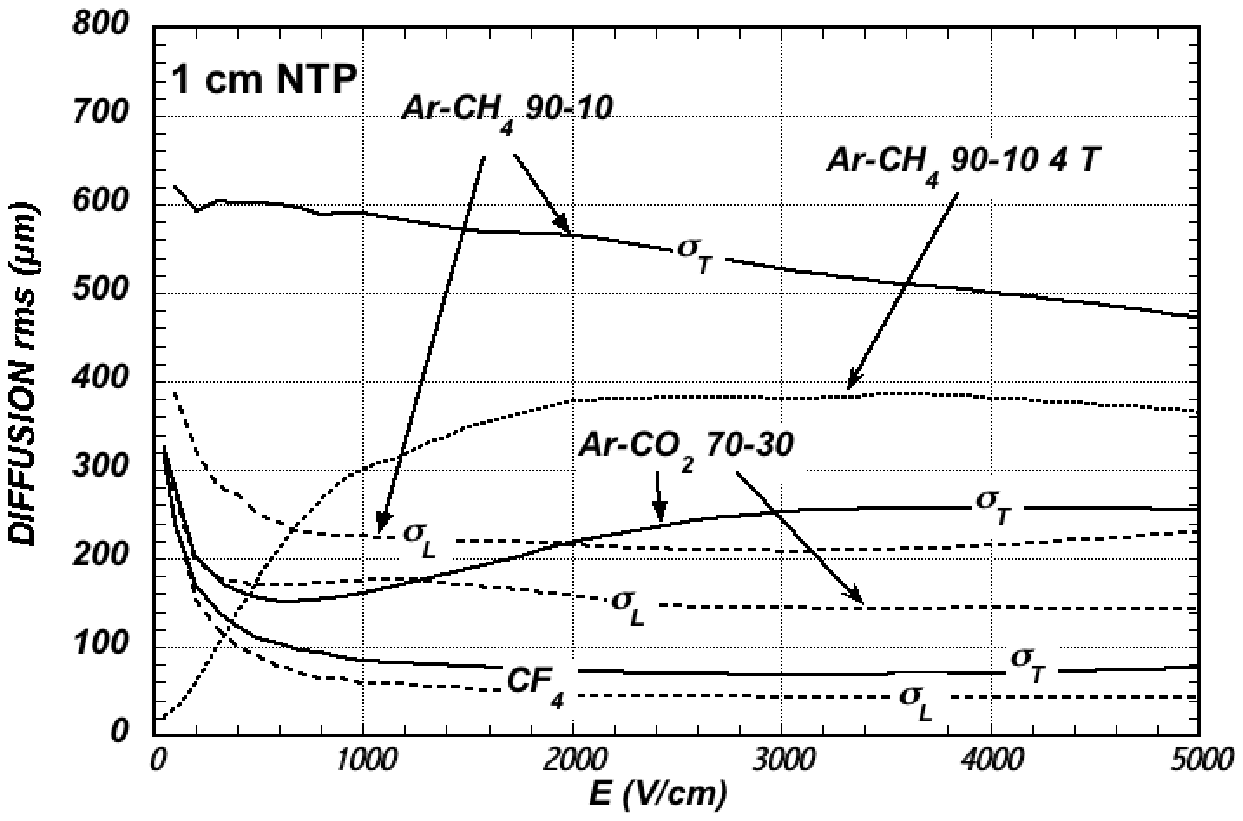}}
 \put(-1.0,52.0){ a) }
 \put(78.0,52.0){ b) }
 \end{picture}
\caption{ a) Computed electron drift velocity with the MAGBOLTZ program
as a function of electric field in several gases at NTP and B=0.
b) Electron longitudinal diffusion ($\sigma_L$) (dashed lines) and
transverse diffusion ($\sigma_T$) (full lines) for 1~cm of drift at NTP
and B~=~0. The dotted line shows $\sigma_T$ for the $Ar/CH_4$(90:10) gas at 4T.}
\label{transverse_parameters}
\end{figure}

 In mixtures containing electronegative molecules such as $O_2$, $H_2 O$ or
CF$_4$, electrons can be captured to form negative ions. Capture
cross-sections are strongly energy-dependent, and therefore the capture
probability is a function of applied field.
For example, the electron is attached to the oxygen molecule
at energies below 1~eV, while dissociative electron attachment to the $CF_4$
occurs mainly in the 6 to 8 eV range~\cite{cf4_summary}.
The three-body electron attachment coefficients may differ greatly for the same addition
in different mixtures.
As an example, at moderate fields (up to 1 kV/cm) the addition of 0.1$\%$ of
oxygen to an Ar/CO$_2$ mixture results in an electron capture probability
about twenty times larger than the same addition to Ar/CH$_4$.

The primary ionization signal is very feeble in a gas layer: 
in one $cm$ of $Ar/CO_2$ (70:30) at NTP around $\sim 100$ electron-ion pairs 
are created (see Eq.~\ref{number_np_nt}).
 Therefore, one has to use ``internal gas amplification'' mechanism 
to generate detectable signal in gas counters; excitation and
subsequent photon emission participate in the avalanche spread processes
and can be also detected by optical means.
If the electric field is increased sufficiently, electrons gain enough energy
between collisions to undergo inelastic collisions with gas molecules.
Above a gas-dependent threshold, the mean free path for ionization,
$\lambda_i$, decreases exponentially with the field; its inverse,
$\alpha=1/\lambda_i$, is the first Townsend coefficient.
In single wire proportional counter, most of the
increase of avalanche particles occurs very close to the anode wires, and a simple electrostatic
consideration shows that the largest fraction of the detected signal is due
to the motion of positive ions receding from the wires.
The electron component, although very fast, contributes very little to the signal.
This determines the characteristic
shape of the detected signals in the proportional mode: a fast rise
followed by a gradual increase.
The slow component, the so-called "ion tail" that limits
the time resolution of the counter, is usually removed by differentiation
of the signal. In uniform fields, $N_0$ initial electrons multiply over a
length $x$ forming an electron avalanche of size $N=N_0\, e^{\alpha x}$; 
$N/N_0$ is the gain of the counter.
With present-day electronics,
proportional gains around $5\cdot 10^3 - 10^4$ are sufficient for detection of minimum
ionizing particles, and noble gases with moderate amounts of polyatomic
gases, such as methane or carbon dioxide, are used.

Positive ions released by the primary ionization or produced in the avalanches drift and
diffuse under the influence of the electric field.
Negative ions may also be produced by electron attachment to gas molecules.
The drift velocity of ions in
the fields encountered in gaseous counters (up to few kV/cm) is typically about
three orders of magnitude lower than for electrons.
The ion mobility, $\mu$, the ratio of drift velocity to electric field, is constant for a given
ion type up to very high fields~\cite{McDaniel73,Shultz77}.
For mixtures, due to a very
effective charge transfer mechanism, only ions with the lowest ionization
potential survive after a short path in the gas. 
The diffusion of ions, both $\sigma_L$ and $\sigma_T$, are proportional to the
square root of the drift time, with a
coefficient that depends on temperature but not on the ion mass.
Accumulation of ions in the gas volume may induce gain reduction and
field distortions.

\section{The Multi-Wire Proportional, Drift and Time Projection Chambers}

 The invention of the Multi-Wire Proportional Chambers (MWPC) revolutionized the
field of radiation detectors~\cite{charpak_1968,annrev_1984_34_285}. 
 In the original design, the MWPC consists of a set of parallel,
evenly spaced, anode wires stretched between two cathode planes.
 Applying potential difference between anodes and cathodes,
field lines and equipotentials develop as shown in Fig.~\ref{MWPCandDrift}a.
 Electrons released in the gas volume drift towards the anodes and produce
avalanche in the increasing field.
 Typical values for the anode wire spacing range between 1 and 5~$mm$, 
the anode to cathode distance is 5 to 10~$mm$.
 The operation gets increasingly difficult at smaller wire spacings, 
which prevented taking this direction for obtaining higher spatial resolution.
For example, the electrostatic repulsion for thin ($10~\mu m$) anode wires
causes mechanical instability above a critical wire length, which is less
than 25~$cm$ for 1-$mm$ wire spacings.
 The signal multiplication process, which begins a few radii from the anode,
is over after a fraction of nanosecond, leaving the cloud of positive ions
receding from the anode.
Detection of charge on the anode wires over a predefined threshold provides the
transverse coordinate to the wire with an accuracy comparable to that
of the wire spacing. With a digital readout and $s=1~mm$ wire spacing,
the spatial resolution is limited to: 
$\sigma = \frac{s}{\sqrt{12}} = 300~\mu m$.
The coordinate along each wire
can be obtained by measuring the ratio of collected charge at
the two ends of resistive wires. 
 The slow motion of positive ions produced in the avalanche is responsible for the largest 
fraction of charge signals detected on all surrounding electrodes; 
a measurement of the charge profile induced on segmented cathodes,
allows bi-dimensional localization of the ionizing event.
 This is the so-called center-of-gravity (COG) method, which permits 
to attain the highest localization accuracy in MWPCs.
Due to the statistics
of energy loss and asymmetric ionization clusters, the position accuracy
is $\sim 50~\mu m~rms$ for tracks perpendicular to the wire plane, but
degrades to $\sim 250~\mu m$ at $30^\circ$ to the normal~\cite{charpak_1979}.

\setlength{\unitlength}{1mm}
\begin{figure}[bth]
 \begin{picture}(125,125)
 \put(-30.0,-105.0){\includegraphics{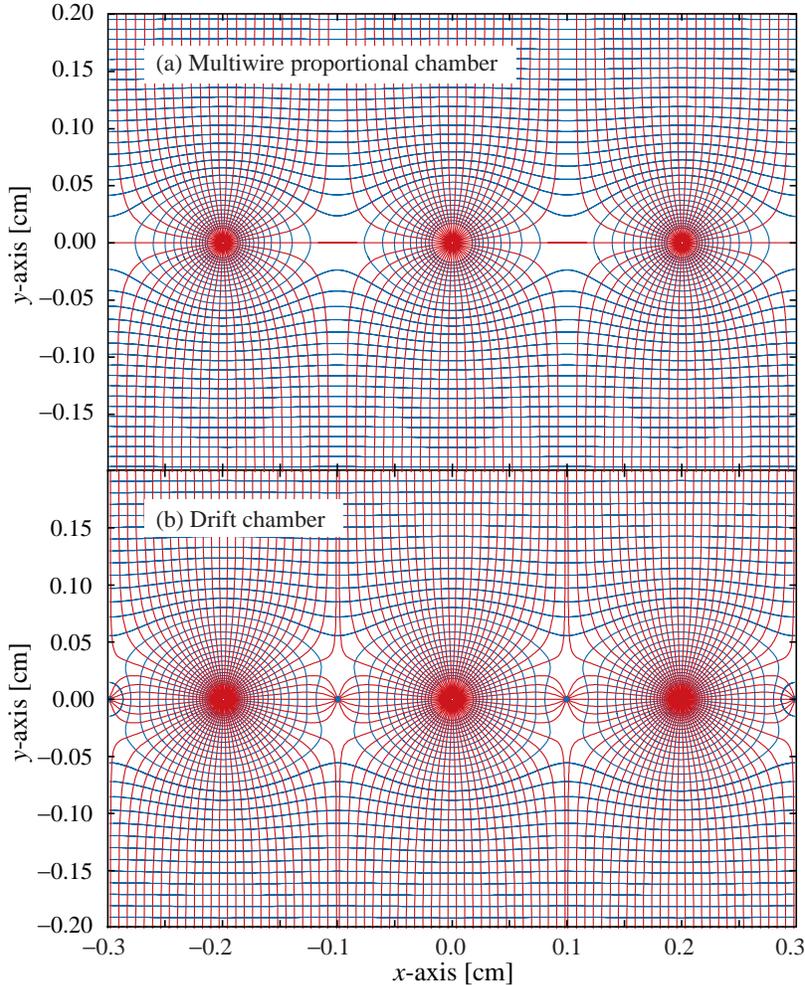}}
 \end{picture}
\caption{ Electric field lines and equipotentials in
(a) a multiwire proportional chamber and
(b) a drift chamber. Thin anodes wires between two cathodes
acts as a set of independent proportional counters.}
\label{MWPCandDrift}
\end{figure}

Drift chambers, developed in the early '70's, can estimate the position
of a track by exploiting the arrival time of electrons at the anodes
if the time of interaction is known\cite{walenta_1971}.
The distance between anode wires is usually several cm allowing coverage
of large areas at reduced cost.
Many designs have been introduced aimed at improving performance.
In the original design, a thicker wire at proper voltage between anodes (field
wire) reduces the field at the middle point between anodes, improving
charge collection (Fig.~\ref{MWPCandDrift}b).
In some drift chambers design, and with the help
of suitable voltages applied to field-shaping electrodes, the electric
field structure is adjusted to improve the linearity of space-to-drift-time
relation, resulting in better spatial resolution~\cite{nima124_1975}.
Drift chambers can reach a spatial resolution from timing measurement
of order 100~$\mu m$ (rms) or
better for minimum ionizing particles, depending on geometry and operating conditions.
A degradation of resolution is observed however for tracks close to the anode wires,
caused by the spread in arrival time of the nearest ionization clusters,
due to primary ionization statistics~\cite{nima156_1978}.
Sampling the drift time on rows of anodes led to the
concept of multiple arrays such as the multi-drift
module~\cite{Bouclier88} and the JET chamber~\cite{Drumm80}.
A measurement of drift time, 
together with the recording of charge sharing from the two ends of the anode wires
provides the coordinates of segments of tracks; the total charge gives information
on the differential energy loss and is exploited for particle identification.

 The ``ultimate'' drift chamber is the TPC concept invented in the 1976~\cite{Nygren78},
which combines a measurement of drift time and charge
induction on cathodes to obtain excellent tracking for high multiplicity
topologies occurring at moderate rates.
It has been the prime choice for large tracking systems in $e^+ e^-$ colliders 
(PEP-4~\cite{pep_proposal_004}, ALEPH~\cite{aleph_tpc}, DELPHI~\cite{delphi_tpc})
and proved its unique resolving power in the heavy ion collisions 
(NA49~\cite{na49_tpc}, STAR~\cite{star_tpc}).
 A TPC consists of a large gas volume, with an uniform electric field applied between
the central electrode and a grid at the opposite side.
 The ionization trails produced by charged particles drift towards the readout end-plate,
where a 2D image of tracks is reconstructed;
the third coordinate is measured using the drift time information.
 In all cases, a good knowledge of electron drift velocity and diffusion properties is
required. This has to be combined with the 3D modeling of electric
fields in the structures, computed with commercial or custom-developed
software~\cite{SIMULATION,MAXWELL}. 
 Conventional readout structure, based on MWPC and pads, is a benchmark
for the ``most modern'' ALICE TPC, designed to cope with extreme
instantaneous particle densities produced in heavy ion collisions
at the LHC.
 This detector incorporates innovative and state of the art technologies,
from the mechanical structures to the readout electronics and data processing
chain~\cite{nima565_2006}.
 To limit distortions to the ALICE TPC intrinsic spatial resolution
($\sigma_{r\phi}\sim$1000~$\mu m$ for 250~cm drift length),
the temperature gradient in a 88~$m^3$ gas volume space
must not exceed 0.1$^{\circ} C$~\cite{meyer}.
For an overview of detectors exploiting the drift time for coordinate measurement see
Ref.~\cite{Grupen_1996} and Ref.~\cite{Blum_2008}.

  Despite various improvements, position-sensitive detectors based on wire structures are
limited by basic diffusion processes and space charge
effects in the gas to localization accuracies around 100~$\mu m$\cite{Aleska00}.
 The presence of slow moving positive ions from electron avalanches
generates a positive space charge in the drift tube,
that modifies electric field and leads to the uncertainty in the space-to-drift time relation.
 In standard operating conditions, the gain of a MWPC starts to drop at particle
rates above $10^4~mm^{-2} s^{-1}$, leading to a loss of detection 
efficiency~\cite{nima124_1975} (see Fig.~\ref{MWPC_MSGC}).
 Together with the practical difficulty to manufacture detectors with sub-$mm$
wire spacing, this has motivated the development of new generation gaseous detectors
for high luminosity accelerators.

\setlength{\unitlength}{1mm}
\begin{figure}[bth]
 \begin{picture}(50,50)
 \put(35.0,-6.0){\includegraphics{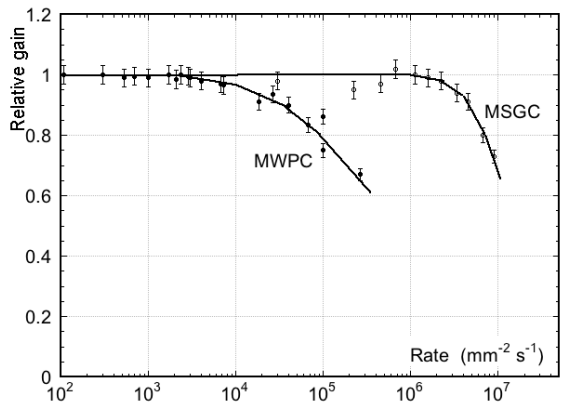}}
 \end{picture}
\caption{ Normalized gas gain as a function of particle rate for
Multi-Wire Proportional Chamber (MWPC)
and Micro-Strip Gas Chamber (MSGC).
\label{MWPC_MSGC}}
\end{figure}

\section{Micro-Pattern Gas Detectors (MPGD)}

 Modern photo-lithographic technology led to the development of
novel Micro-Pattern Gas Detector (MPGD) concepts~\cite{annrev_1984_34_285},
revolutionizing cell size limitations for many gas detector applications.
 By using pitch size of a few hundred microns,
an order of magnitude improvement in granularity over wire chambers,
these detectors offer intrinsic high rate capability ($>10^6$ Hz/$mm^2$), excellent
spatial resolution ($\sim30~\mu$m), multi-particle resolution ($\sim500~\mu$m),
and single photo-electron time resolution in the ns range.

 The Micro-Strip Gas Chamber (MSGC), a concept invented in 1988,
was the first of the micro-structure gas detectors\cite{msgc_1988}.
It consists of a set of tiny parallel metal strips
laid on a thin resistive support,
alternatively connected as anodes and cathodes (see Fig.~\ref{MSGC_sketch_photo}a).
The principle of MSGC resembles a multi-anode proportional counter, with fine printed
strips instead of wires, see Fig.~\ref{MSGC_sketch_photo}b.
 Owing to the small anode-to-cathode distance ($\sim 100~\mu m$), the fast collection
of positive ions by nearby cathode strips reduces space charge build-up, and 
pushes the maximum rate capability of MSGC, compared to MWPC,
by more than two orders of magnitude~\cite{barr_1998} (see Fig.~\ref{MWPC_MSGC}).
Despite their promising performance, experience with MSGCs has raised serious concerns about
their long-term behavior.
There are several major processes, particularly at high rates,
leading to the MSGC operating instabilities: 
substrate charging-up and time-dependent modification of electric field,
surface deposition of polymers (aging) and 
destructive micro-discharges under exposure to heavily ionizing 
particles~\cite{annrev_1984_34_285,bouclier_nima381}.
 The problem of discharges is the intrinsic limitation of all single-stage micro-pattern
detectors in hadronic beams~\cite{peskov_nima397_243,ivan_nima422_300,bressan_nima424}.
 Whenever the total charge in the avalanche exceeds a value of $10^7 - 10^8$ electron-ion
pairs (Raether limit), an enhancement of the electric field in front of and 
behind the primary avalanche induces the fast growth of a filament-like streamer
followed by breakdown.
 This has been confirmed under a wide range of operating 
conditions and multiplying 
gaps~\cite{ivan_ieeetns_45_258,fonte_ieeetns_46_321,iacobaeus_ieeetns_49_1622,arXiv_09110463}.
In the high fields and narrow gaps, the MSGC turned out to be prone to irreversible discharges induced 
by heavily ionizing particles and destroying the fragile 
electrode structure~\cite{bagaturia_2002} (see Fig.~\ref{MSGC_sketch_photo}c).
 Nevertheless, the detailed studies on their properties, and in particular,
on the radiation-induced processes leading to discharge breakdown, led
to the development of mature technologies and novel approaches
(GEM and Micromegas)
with similar performances, improved reliability and radiation hardness.
 The MPGD structures can be grouped in two large families: micromesh-based 
detectors and hole-type structures.
The micromesh-based structures include: Micromegas, ``Bulk'' Micromegas, 
``Microbulk'' Micromegas and ``InGrid''. The Hole-type structures are: 
GEMs, THGEM, RETGEM and Micro-Hole and Strip Plate (MHSP) elements. 

\setlength{\unitlength}{1mm}
\begin{figure}[bth]
 \begin{picture}(40,40)
 \put(1.0,-6.0){\includegraphics{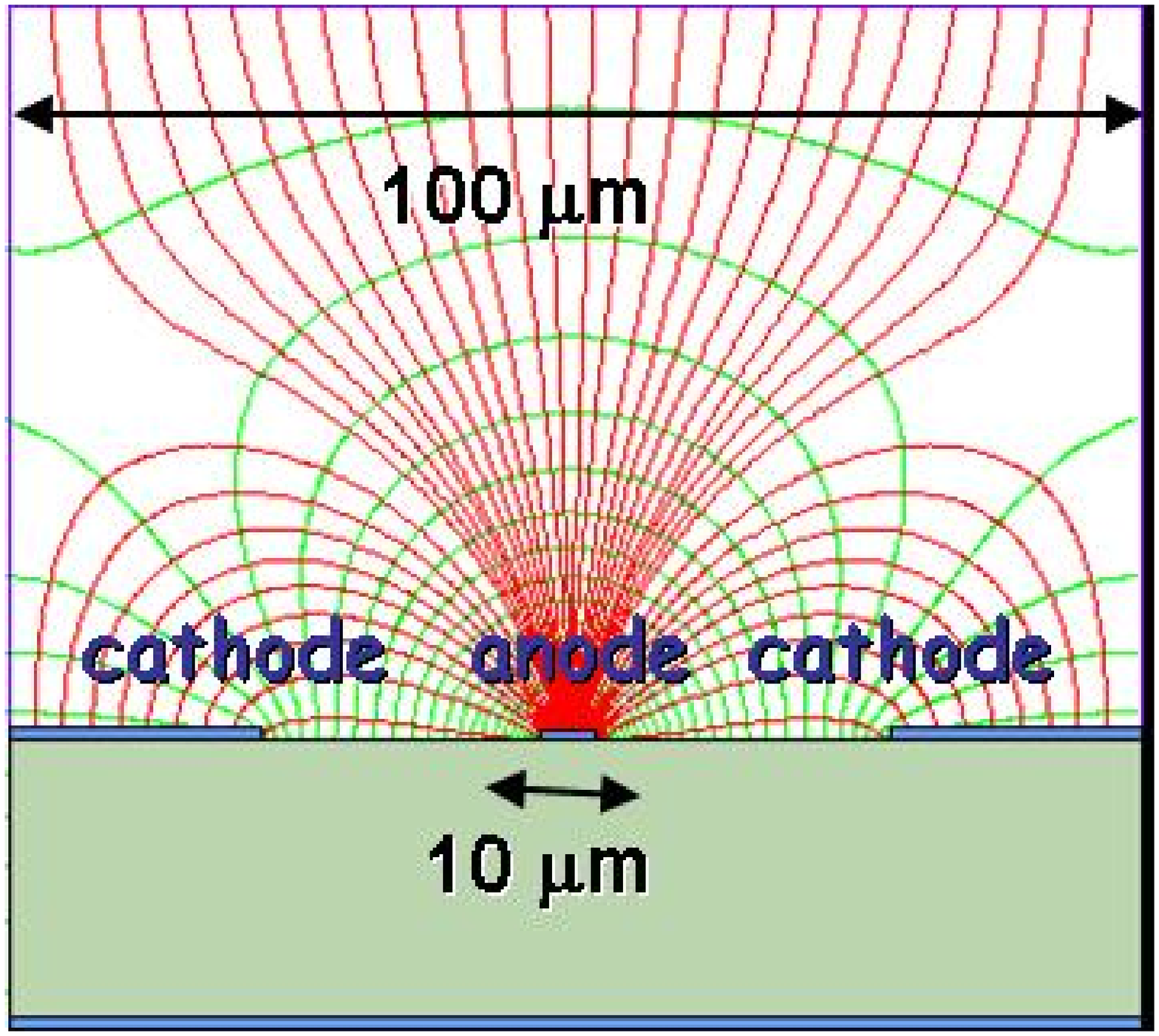}}
 \put(55.0,-5.0){\includegraphics{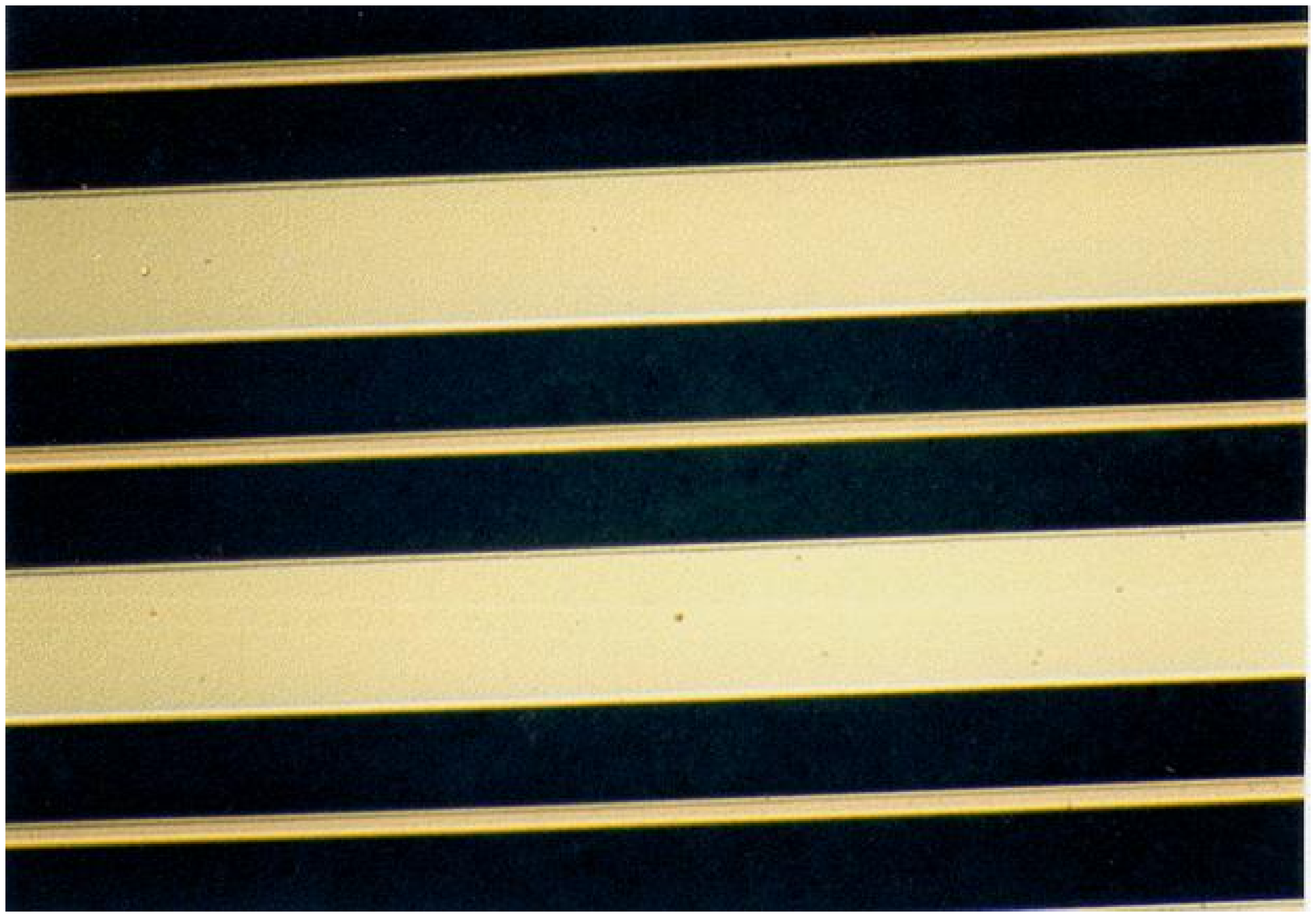}}
 \put(108.0,-5.0){\includegraphics{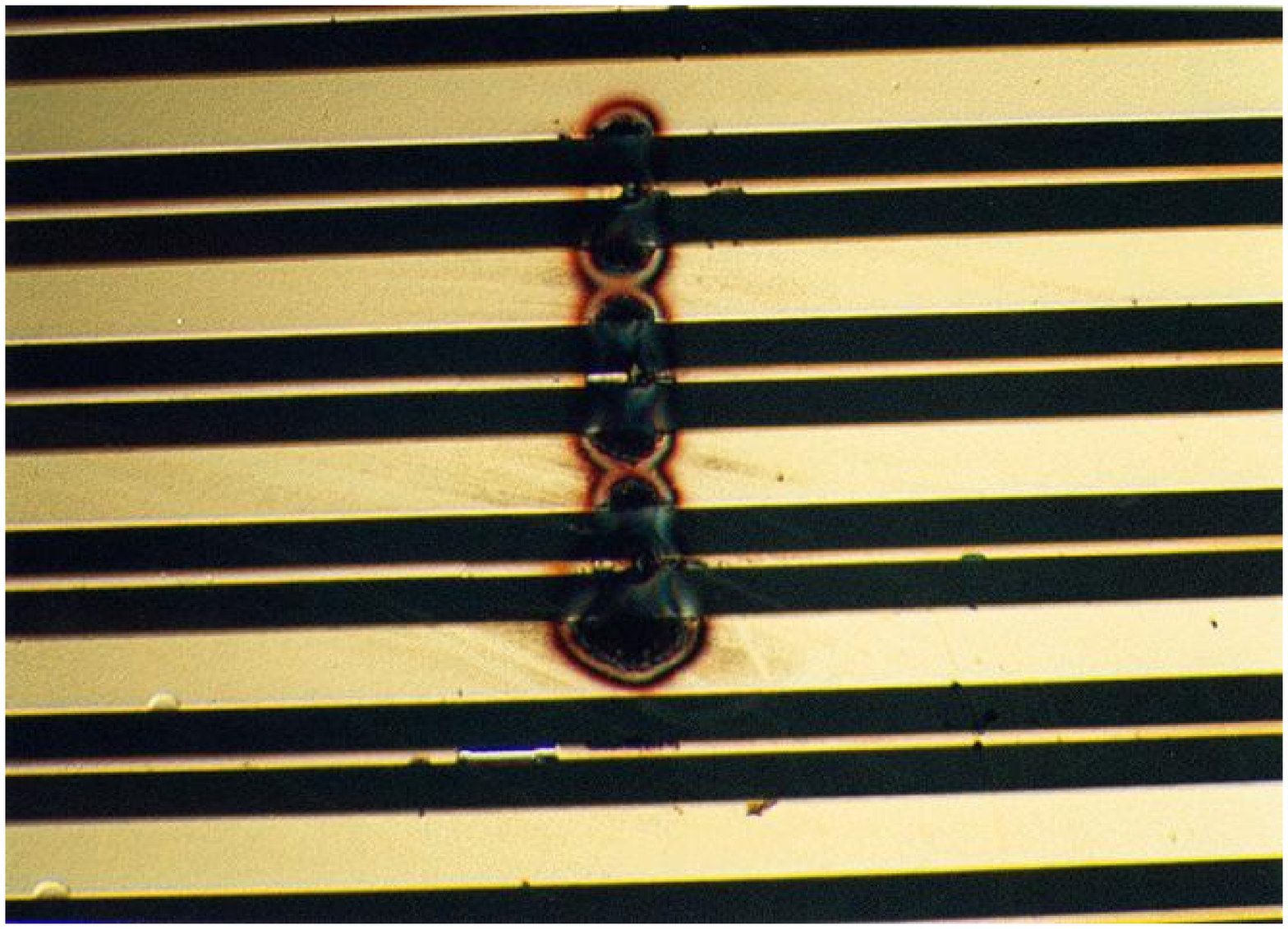}}
 \put(-1.0,37.0){ a) }
 \put(51.0,37.0){ b) }
 \put(105.0,37.0){ c) }
 \end{picture}
\caption{ a) Schematic view, equipotentials and field lines in the MSGC; b)  
Microscopic image of the MSGC. On an insulating substrate, thin metallic anode
strips alternate with wider cathodes; the pitch is 200~$\mu m$;
c) Image of MSGC electrodes damaged by discharge. The very thin
metal layers of MSGCs (few hundred nanometers)  makes them vulnerable
for discharges, which can easily destroy fragile structure.}
\label{MSGC_sketch_photo}
\end{figure}

\setlength{\unitlength}{1mm}
\begin{figure}[bth]
 \begin{picture}(40,40)
 \put(-18.0,-42.0){\includegraphics{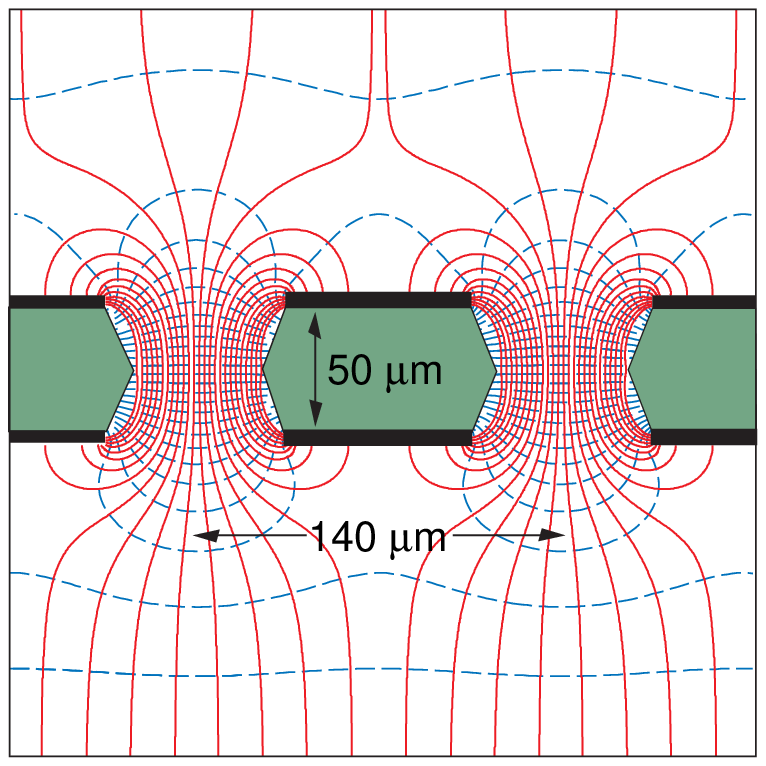}}
 \put(50.0,-5.0){\includegraphics{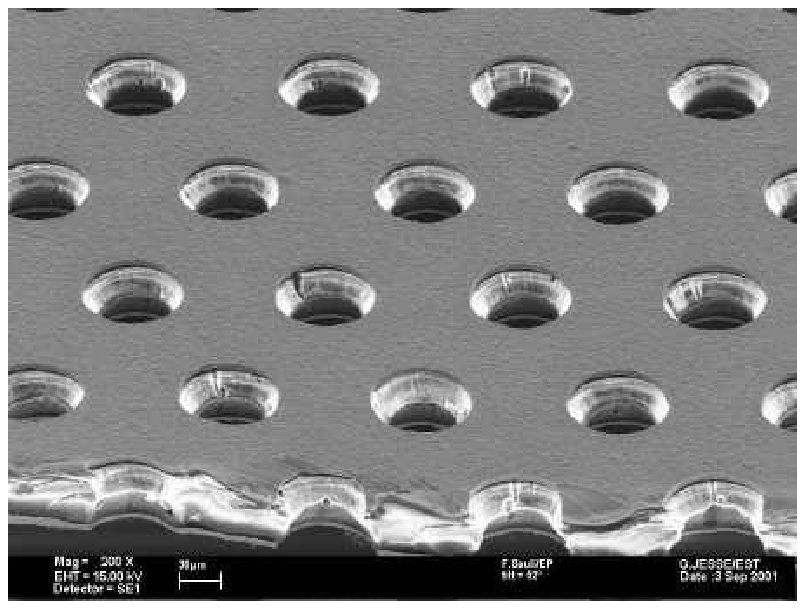}}
 \put(103.0,-8.0){\includegraphics{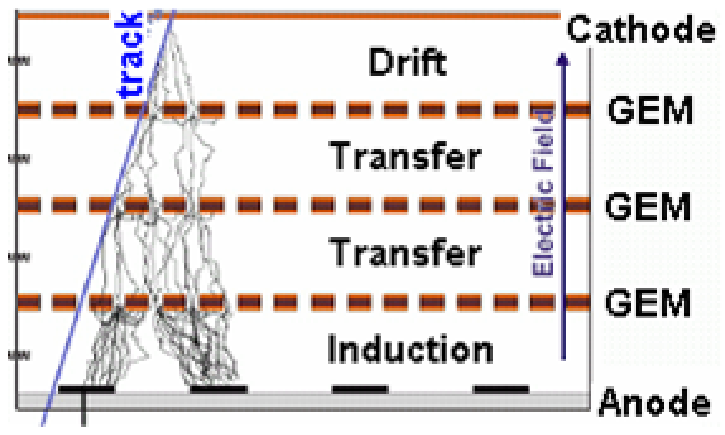}}
 \put(-1.0,37.0){ a) }
 \put(47.0,37.0){ b) }
 \put(102.0,37.0){ c) }
 \end{picture}
\caption{ a) Schematic view and typical dimensions of the hole structure in the GEM
amplification cell. Electric field lines (solid) and equipotentials
(dashed) are shown;
b) Electron microscope view of a GEM foil etched on a copper-clad, 50 $\mu m$ thick polymer foil. 
The hole's diameter and pitch are 70 $\mu m$ and 140 $\mu m$;
c) Schematic view of the triple-GEM detector.}
\label{GEM_sketch_photo}
\end{figure}

Introduced in 1996 by F. Sauli~\cite{GEM_NIMA386}, a Gas Electron Multiplier (GEM) 
detector consists of a thin-foil copper-insulator-copper sandwich chemically 
perforated to obtain a high density of holes.
The GEM manufacturing method, developed at CERN, is a refinement of the double-side
printed circuit technology.
The hole diameter is typically between 25~$\mu$m and 150~$\mu$m,
while the corresponding distance between holes varies between 50~$\mu$m and 200~$\mu$m.
 The central insulator is usually (in original design)
the polymer kapton, with a thickness of 50~$\mu$m (see Fig.~\ref{GEM_sketch_photo}b).
 Controlled etching of GEM foils (decreasing the thickness of the copper layer from 5 to 1 $\mu m$)
allows to reduce material budget in triple-GEM to 1.5$\times$10$^{-3} X_0$,
which is about one half of a 300-$\mu m$-thick Si-microstrip detector~\cite{bondar_nima556}.
Application of a potential difference between the two sides of the GEM 
generates the electric fields indicated in Fig.~\ref{GEM_sketch_photo}a.
Each hole acts as an independent proportional counter.
Electrons released by the primary ionization particle in the upper
conversion region (above the GEM foil) drift into the holes,
where charge multiplication occurs in the high electric field (50~-~70~kV/cm).
Most of avalanche electrons are transferred into the gap below the 
GEM~\cite{bressan_nima425,bachmann_nima438}.
Several GEM foils can be cascaded (see Fig.~\ref{GEM_sketch_photo}c), allowing
the  multi-layer GEM detectors to operate at overall gas gain above $10^4$
in the presence of highly ionizing particles, while strongly
reducing the risk of discharges ($< 10^{-12}$ per hadron)~\cite{bachmann_nima470,bachmann_nima479}.  
This is a major advantage of the GEM technology.
 A unique property of GEM detector is a full decoupling of the amplification stage (GEM)
and the readout electrode (PCB), which operate at unity gain and serves only as a charge collector.
 Localization can then be performed by collecting the charge on a
one- or two-dimensional readout board of arbitrary pattern, placed below the last GEM.
The signal detected on the PCB is entirely due to the electron collection, without
a slow ion tail, and is typically few tens of nanosecond for 1~$mm$-wide induction gap.
Cascaded GEMs reach gains above $10^5$ with single electrons; this permitted conceiving 
gaseous imaging photo-multipliers (GPM) with single photon sensitivity~\cite{breskin_nima513}.
 Moreover, with an appropriate choice of GEM fields and geometry, both photon and ion feedback
can be considerably suppressed.

 The MHSP is a GEM-like hole-electrode with anode- and cathode-strips etched on the bottom 
face of the GEM; the avalanche developed inside the hole is further multiplied on the 
anode strips~\cite{veloso_revsciinst_71_2371}. When operating the MHSP and biasing the strips in reverse mode,
i.e. using an extra electrode to attract positive ions, a breakthrough in ion 
blocking capability has been achieved in MHSP/GEM cascaded multipliers~\cite{jinst2_p08004_2007}.

\setlength{\unitlength}{1mm}
\begin{figure}[bth]
 \begin{picture}(40,40)
 \put(-45.0,-133.0){\includegraphics{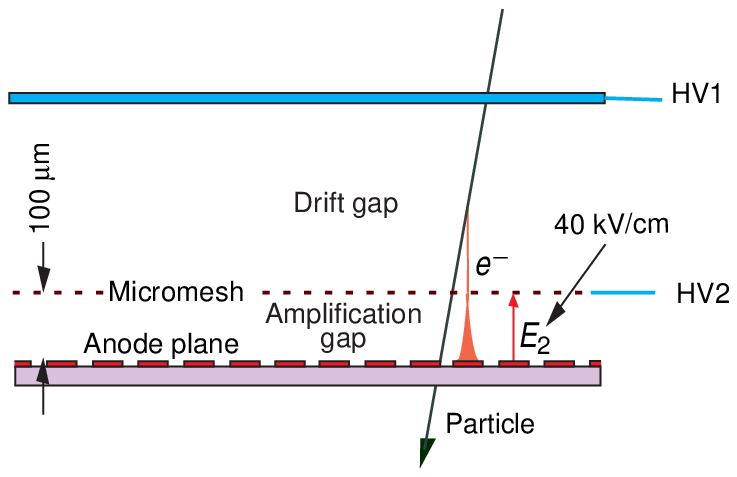}}
 \put(63.0,0.0){\includegraphics{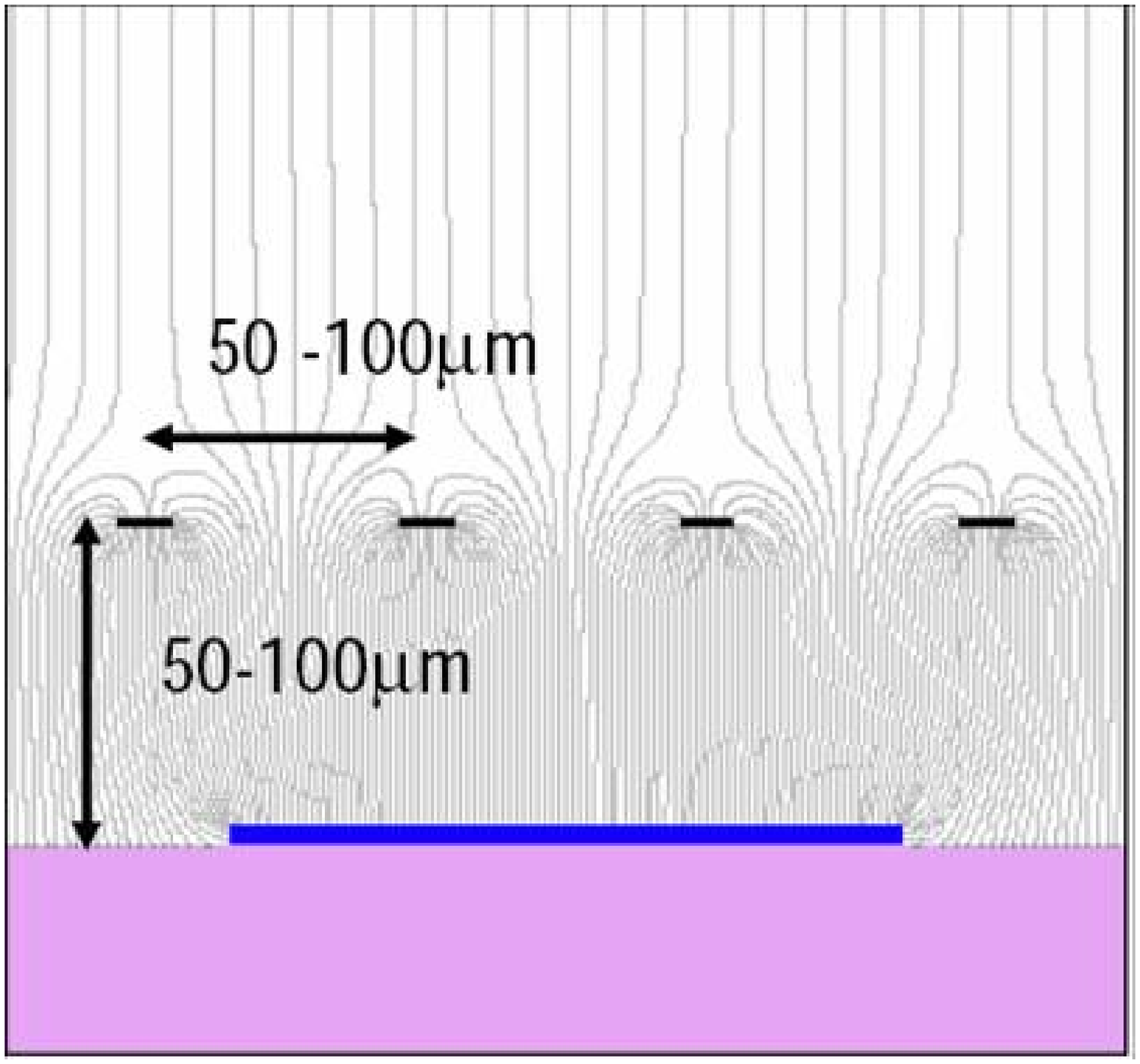}}
 \put(110.0,0.0){\includegraphics{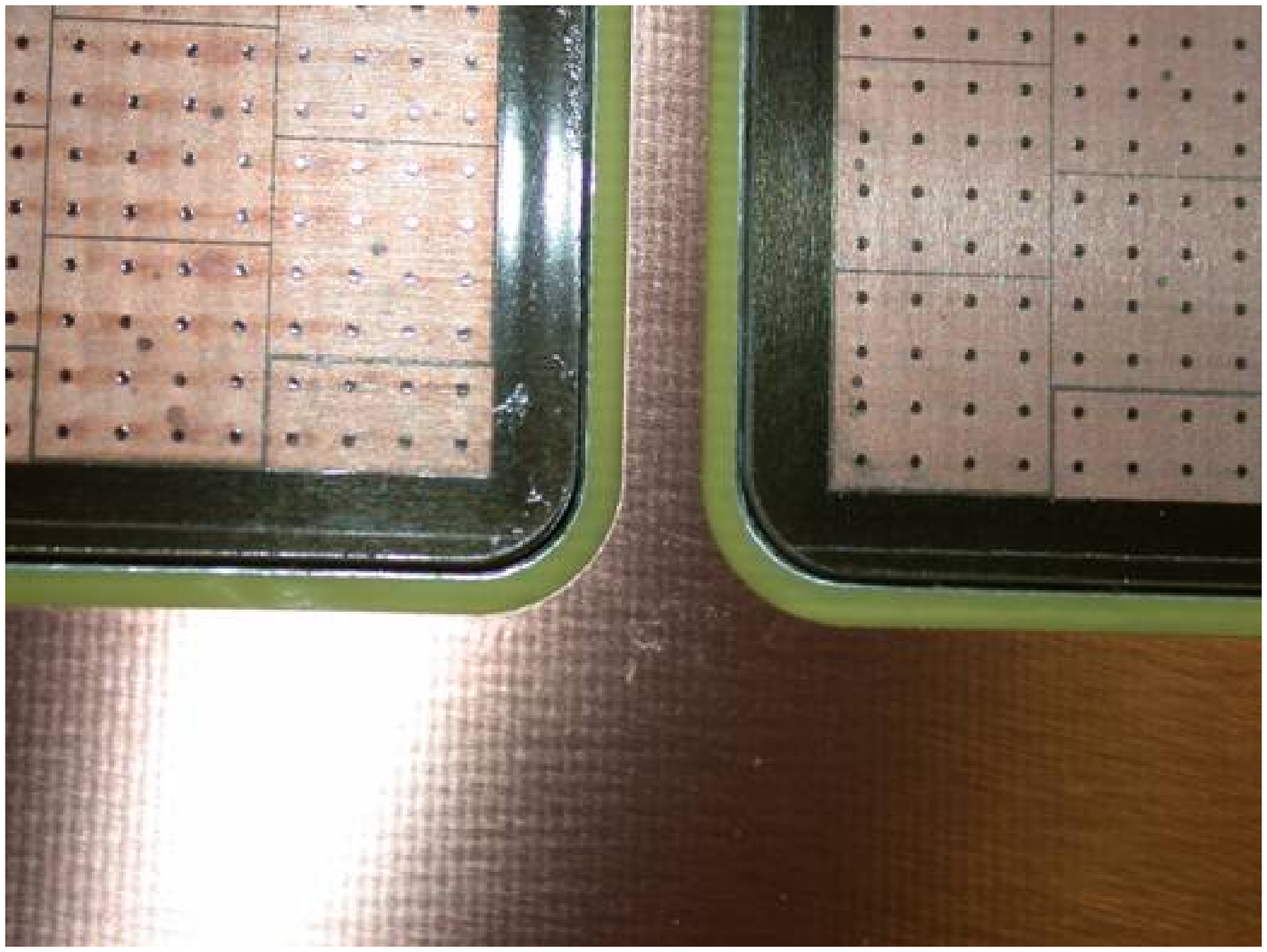}}
 \put(-1.0,37.0){ a) }
 \put(59.0,37.0){ b) }
 \put(107.0,37.0){ c) }
 \end{picture}
\caption{ a) Schematic view of the Micromegas detector (not to scale). A few $mm$
drift gap is between the cathode electrode and micromesh, and the 100~$\mu m$
amplification gap is defined by the micromesh and the anode strip readout;
b) Electric field map in Micromegas. A metallic micromesh separates 
a low-field ($\sim 1~kV/cm$) drift region from the high-field multiplication 
region ($\sim 40-80~kV/cm$).
c) Photograph of the ``Bulk'' Micromegas detectors. Pillars of 400~$\mu m$ diameter
every 2~mm are visible.
}
\label{Micromegas_sketch_photo}
\end{figure}

 Introduced in 1996 by Y. Giomataris, the Micromegas is a thin parallel-plate avalanche counter, 
as shown in Fig.~\ref{Micromegas_sketch_photo}a~\cite{Micromegas_NIMA376}. 
It consists of a few mm drift region and a
narrow multiplication gap (25-150~$\mu$m) between a
thin metal grid (micromesh) and the readout electrode (strips or pads of
conductor printed on an insulator board).
 To preserve a distance between the anode and the micromesh,
regularly spaced pillars from insulating material are used.
  Electrons from the primary ionization drift through the holes
of the mesh into the narrow multiplication gap, where
they are amplified.
The electric field is homogeneous both in the drift
(electric field $\sim$1~kV/cm) and amplification ($\sim$40-80~kV/cm) 
gaps (see Fig.~\ref{Micromegas_sketch_photo}b).
Due to the narrow multiplication region,
small variations of the amplification gap are counteracted by an inverse
variation of the Townsend coefficient,
thus improving the uniformity and stability of response over a large area~\cite{giomataris_nima419_239}.
Positive ions are quickly removed by the micromesh; this prevents 
space-charge accumulation and induces very fast signals with only a small 
ion tail, 50~-~100~$ns$ length.
The Micromegas retains the rate capability and energy resolution 
of the parallel-plate counter.
The small amplification gap is a key element in Micromegas operation, giving rise
to excellent spatial resolution: 
12~$\mu$m accuracy, limited by the micromesh pitch, has been achieved
for MIPs~\cite{derre_nima459_523}, a very good energy resolution
($\sim$11$\%$ FWHM with 8~keV $X$-rays)~\cite{charpak_nima_478_26}
and single photo-electron time resolution better than 1~$ns$\cite{derre_nima449_314}.
 Efforts have been focused on producing the Micromegas amplification region as a single piece
using the newly developed ``Bulk'' method~\cite{giomataris_nima_560_405}.
A woven mesh is laminated on a PCB covered by a photo-imageable polyimide film, and the 
pillars are made by a photochemical technique with insulation through the grid. 
Such a ``all-in-one'' detector, called ``Bulk'' Micromegas, is robust and 
allows the regular production of large, stable and unexpensive detector modules.
Fig.~\ref{Micromegas_sketch_photo}c shows photo of the large ``Bulk'' Micromegas
(36*34 cm$^2$ active area) produced for the T2K TPC~\cite{anvar_nima602_415}.
 A new Micromegas manufacturing technique, based on kapton etching technology,
has been recently developed, resulting in further improvement of the detector
characteristics, such as flexible structure, low mass, high radio-putiry,
uniformity and stability~\cite{microbulk_jinst5_p02001_2010}.
Excellent energy resolution has been obtained, reaching 11~$\%$ FWHM for the 5.9~keV
$X$-rays and 1.8~$\%$ FWHM of the 5.5~$MeV$ $\alpha$-peak for the $Am^{241}$ source.

 The success of GEMs and glass capillary plates triggered the development of coarse 
and more robust structures, ``optimized GEM''~\cite{thgem_nima478_377,thgem_TNS50_809} 
followed by THGEM~\cite{thgem_nima535_303,thgem_nima598_107,thgem_nima558_475} 
gaseous multiplier. 
These are produced by standard printed circuit technology: mechanical drilling of 
0.3-1~$mm$ diameter holes, etched at their rims to enhance high-voltage stability 
(see Fig.~\ref{THGEM_RETGEM}a and~\ref{THGEM_RETGEM}b); different PCB materials can be used, of typical
thicknesses of 0.4-1~$mm$ and hole spacing of 0.7-1.2~$mm$. 
These electron multipliers exhibit specific features: the geometrical parameters 
can be scaled from GEM ones, but the microscopic behavior of the electrons, 
in particular diffusion in the gas, does not scale.
 The electron collection and transport between cascaded elements
is more effective than in GEM because the
THGEM's hole diameter is larger than the electron's diffusion range  
when approaching the hole.
The THGEM detectors
have been studied in laboratory, 
also in gaseous-photomultiplier 
configurations - coupled to semitransparent or reflecting CsI 
photo-cathodes (see Fig.~\ref{THGEM_RETGEM}a)~\cite{thgem_nima553_35}.
Effective gas amplification factors of $10^5$ and $10^7$ 
can be reached in single and cascaded double-THGEM elements,
which permits efficient imaging of light at 
single-photon level~\cite{thgem_nima558_475}.
Stable operation 
at photon fluxes exceeding $1~MHz/mm^2$ was recorded
together with sub-$mm$
localization accuracy and timing in the 10~$ns$ 
range~\cite{thgem_nima598_107,thgem_jinst5_p01002}.

\setlength{\unitlength}{1mm}
\begin{figure}[bth]
 \begin{picture}(40,40)
\put(-3.0,-8.0){\includegraphics{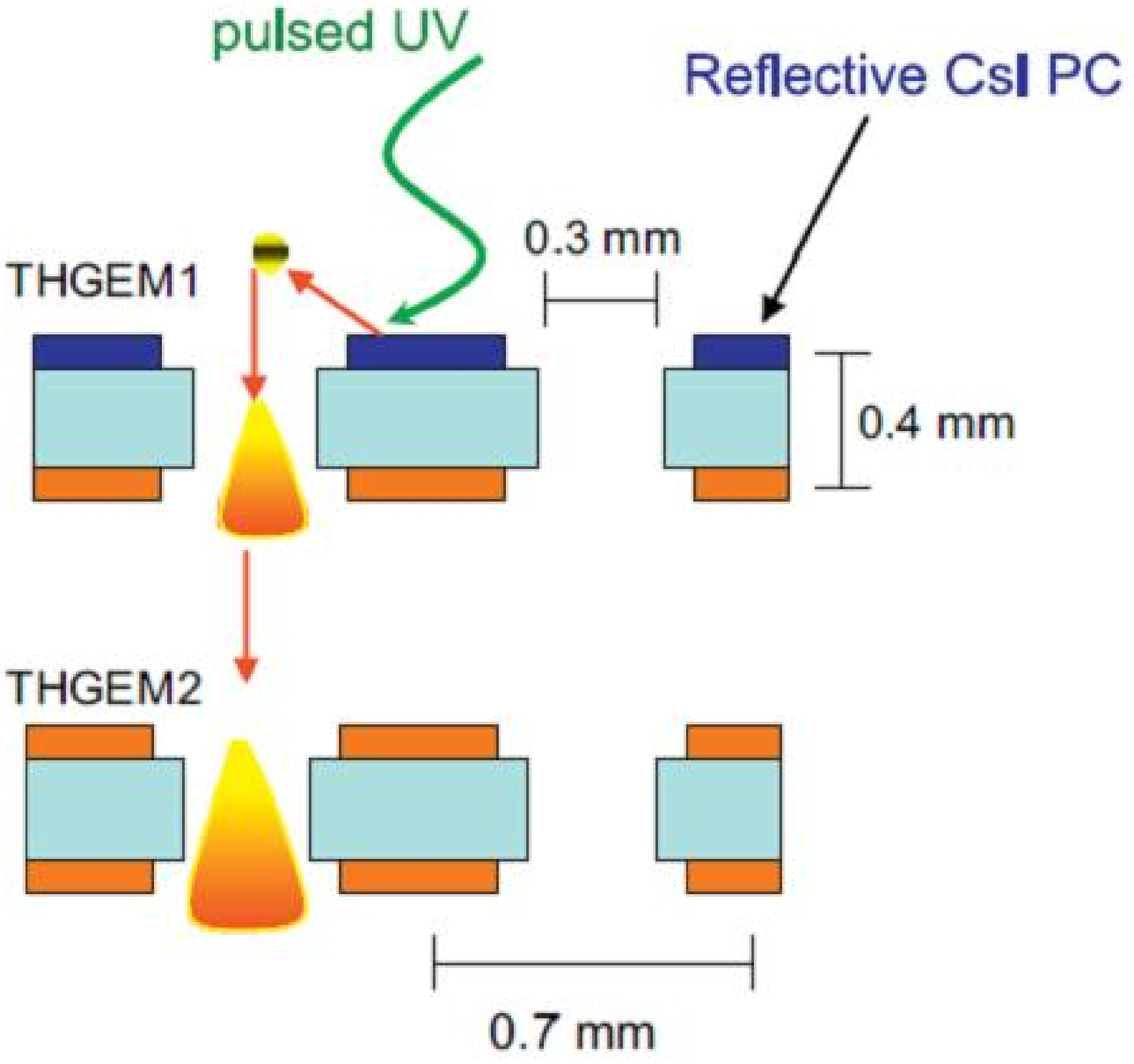}}
 \put(53.0,-5.0){\includegraphics{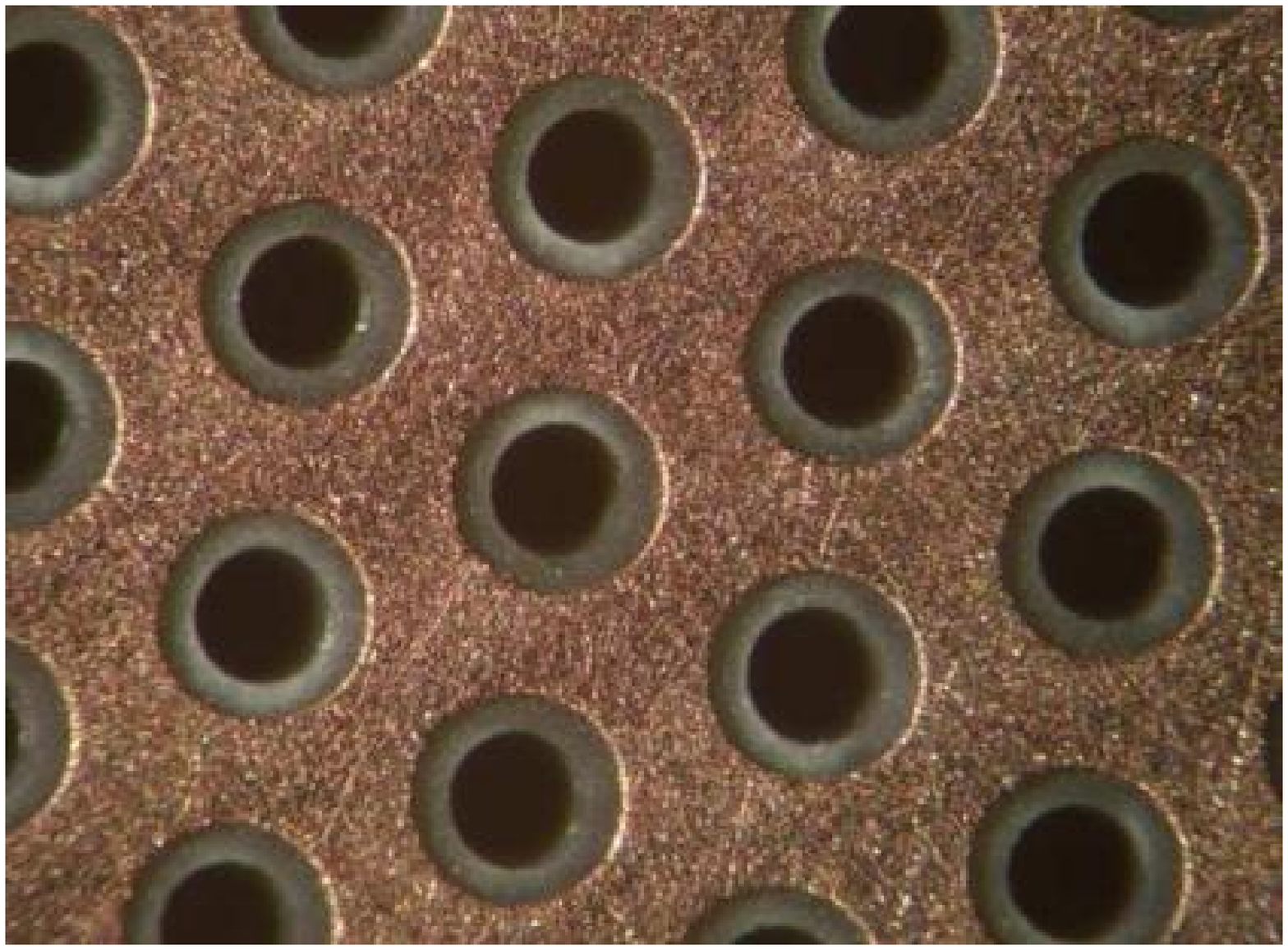}}
 \put(110.0,-5.0){\includegraphics{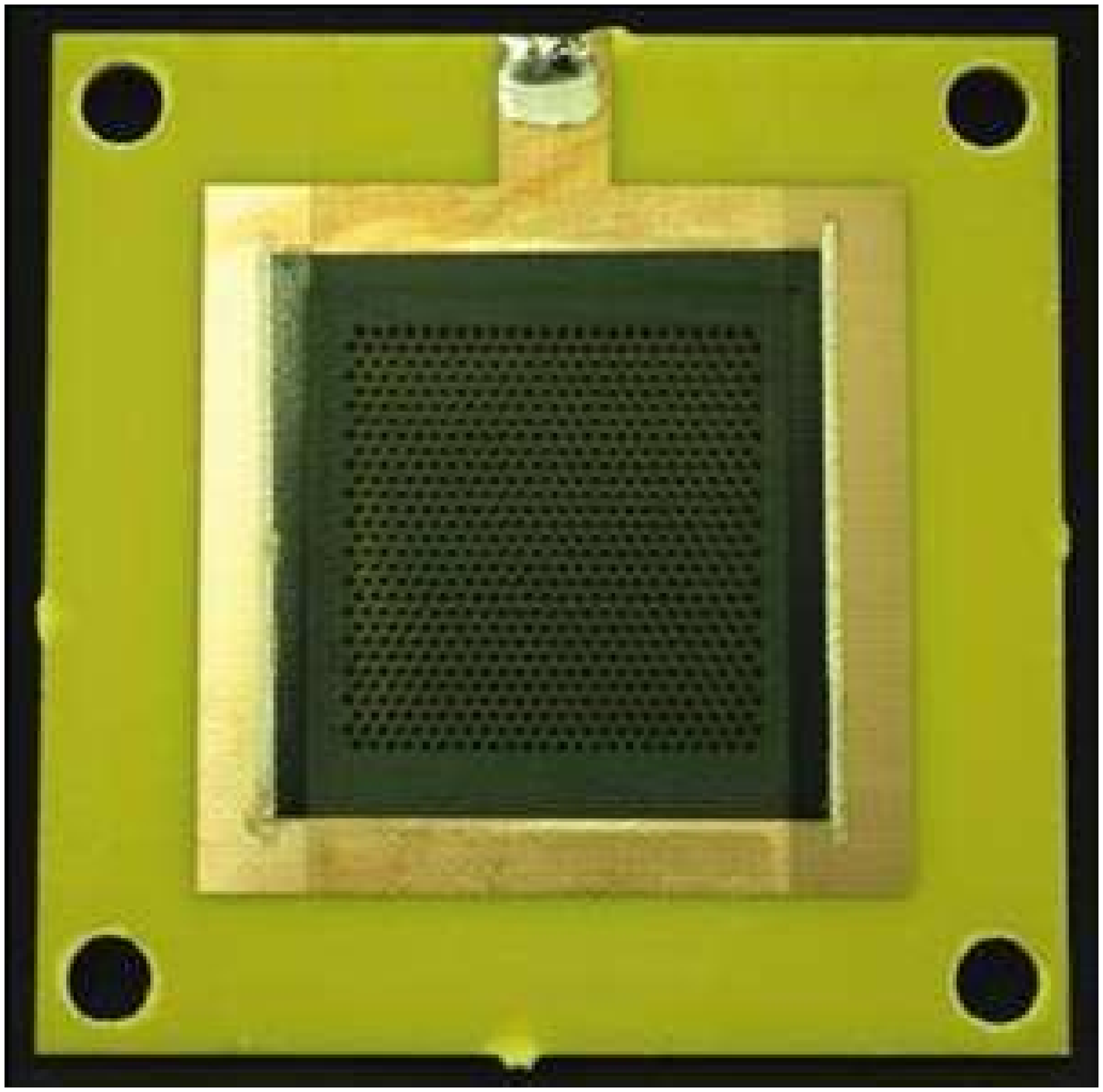}}
 \put(-1.0,37.0){ a) }
 \put(50.0,37.0){ b) }
 \put(106.0,37.0){ c) }
 \end{picture}
\caption{ 
a) Schematic view of a double-THGEM with a reflective $CsI$ PC deposited
on top of first THGEM; b) Photo of the Thick GEM (THGEM) multiplier. 
A rim of 0.1~$mm$ is chemically etched around the mechanically drilled holes 
to reduce discharges; 
c) Photo of the RETGEM detector with resistive kapton electrodes.
}
\label{THGEM_RETGEM}
\end{figure}

A novel spark-protected version of thick GEM with resistive electrodes (RETGEM) 
has been recently developed, where the Cu-clad conductive electrodes are replaced by resistive 
materials~\cite{oliveira_nima_567_362,dimauro_nima_581_225,peskov_nima_610_169}.
Sheets of carbon loaded kapton 50 $\mu m$ thick or screen-printed resistive surface are 
attached onto both surfaces of the PCB to form resistive-electrode structure; holes 0.3~$mm$ 
in diameter with a pitch of 0.6~$mm$ are mechanically drilled (see Fig.~\ref{THGEM_RETGEM}b).
 At low counting rates, the detector operates as a conventional THGEM with metallic 
electrodes, while at high intensities and in case of discharges the behavior is similar 
to a resistive-plate chamber.

\section{Micro-pattern Gas Detector Applications}

 The performance and robustness of MPGDs have encouraged their use
in high-energy and nuclear physics, UV and visible photon detection,
astroparticle and neutrino physics, $X$-ray imaging and neutron detection and medical physics.
Common themes for future applications are low mass, large active areas, 
high spatial resolution, high rate capabilities and high radiation tolerance. 

Due to the wide variety of geometries and flexible operating parameters, MPGDs are 
a common choice for tracking and triggering detectors in nuclear- and particle-physics.
 COMPASS, a high-luminosity experiment at CERN,
pioneered the use of large-area ($\sim40\times40$~cm$^2$) GEM and
Micromegas detectors close to the beam line with particle rates of 25~kHz/mm$^2$.
Both technologies achieved a
tracking efficiency of close to 100$\%$ at gas gains of about $10^4$, a
spatial resolution of 70--100~$\mu$m and a time resolution of
$\sim10$~ns~\cite{gemcompass,micromegascompass}. 
For the future COMPASS physics program, a set of Micromegas detectors
with pixelized readout in the central region is being proposed~\cite{neyret_jinst4_p12004}.
High resolution planar triple-GEM detectors are used in the LHCb Muon System~\cite{lhcb},
for the TOTEM Telescopes~\cite{lami_2009IEEE} and also being developed for
PANDA experiment at the future FAIR facility~\cite{2009IEEE_vanderbroucke}, 
SBS spectrometer for Hall A at Jefferson lab~\cite{sbs_spectrometer},
and as a forward Tracker of STAR experiment at RHIC~\cite{2009IEEE_simon}. 
Using fast $CF_4$-based mixtures, a time resolution of about 5~ns $rms$
is achieved~\cite{barouch}, adequate to resolve two bunch crossings
at the Large Hadron Collider (LHC).
Therefore, large area MPGDs are currently being considered for the LHC machine upgrade program (sLHC),
aiming to increase the luminosity by a factor of $\sim 10$,
as a potential replacement of conventional gaseous detectors.
Micromegas coupled to the CMOS readout may be employed as 
new vertex detector system at the sLHC~\cite{rd51note_2009_006}.
 GEMs and Micromegas can be also bent to form cylindrically curved 
ultra-light detectors, as preferred for inner tracker (barrel) 
and vertex applications~\cite{bencivenni_2009IEEE,aune_nima_604_53}. 

For the future International Linear Collider applications, both 
GEM and Micromegas devices are foreseen as one of the main options for 
the TPC~\cite{2009IEEE_lentdecker,arogancia_nima602_403,matsuda_jinst5_p01010}.
Compared to wire chambers, they offer number of advantages:
negligible E$\times$B track distortion effects, the narrow pad 
response function (PRF) and the intrinsic suppression of ion back-flow,
relaxing the requirements on gating of the devices and, depending on the 
design, possibly allowing non-gated operation of  TPCs.
Large-area MPGD systems are also being studied as a potential solution 
for highly granular digital hadron calorimeter.
Implementations in GEM~\cite{2009IEEE_jaeyu}, Micromegas~\cite{adloff_jinst4_11023}, 
and THGEM technologies have been proposed.

 In the recent years there has been a considerable progress in the field of
gaseous photomultipliers by combining MPGD with semi-transparent
or reflective $CsI$-photocathodes (PC) to localize single 
photoelectrons~\cite{chechik_nima502_195,chechik_nima595_116,buzulutskov_physpart_39_424}.
 The operation of MPGD-based photomultipliers in $CF_4$ with $CsI$-PC
could form the basis of new generation windowless Cherenkov detectors,
where both the radiator and the photosensor operate in the same gas.
Exploiting this principle, originally proposed for 
Parallel Plate Avalanche Chamber~\cite{gionamatis_nima310_585},
a Hadron Blind Detector 
was successfully operated using triple-GEM amplification system
with $CsI$-PC for the PHENIX experiment at RHIC~\cite{fraenkel_nima546_466,2009IEEE_woody}.
 A hadron-blindness property is achieved by reversing the direction of the drift field
$E_D$, therefore pushing primary ionization produced by charged particles towards the mesh.
In this configuration photoelectrons released from the $CsI$-PC surface are still
effectively collected into the GEM holes due to the strong electric field inside holes.
For many applications in photon detectors, a rather coarse (sub-$mm$)
spatial resolution is usually sufficient. Therefore THGEM and RETGEM
are under study for RICH detector upgrades for COMPASS and ALICE~\cite{mauro_ieeetns_56_1550}.

\setlength{\unitlength}{1mm}
\begin{figure}[bth]
 \begin{picture}(45,45)
 \put(0.0,-6.0){\includegraphics{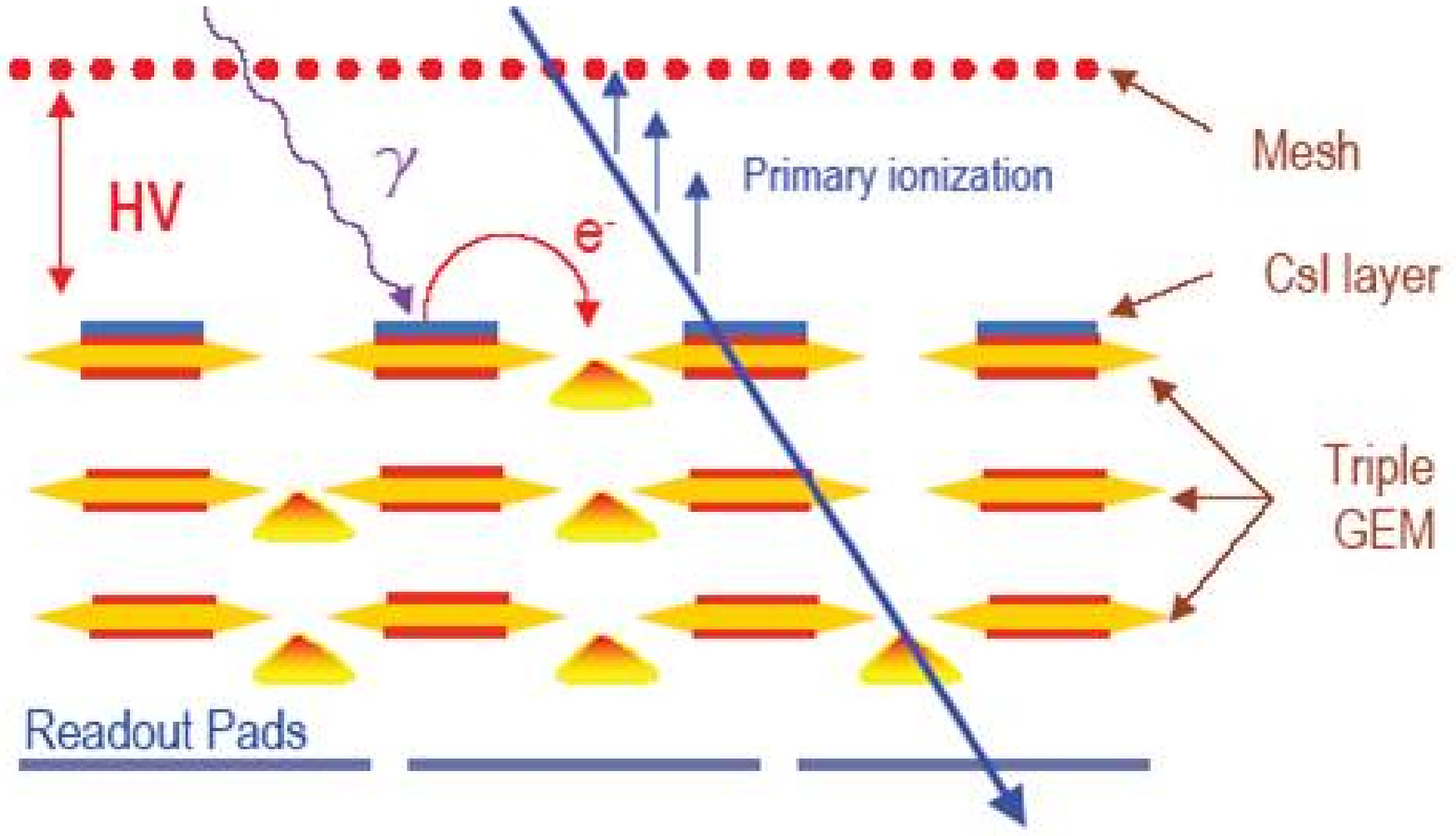}}
 \put(80.0,-7.0){\includegraphics{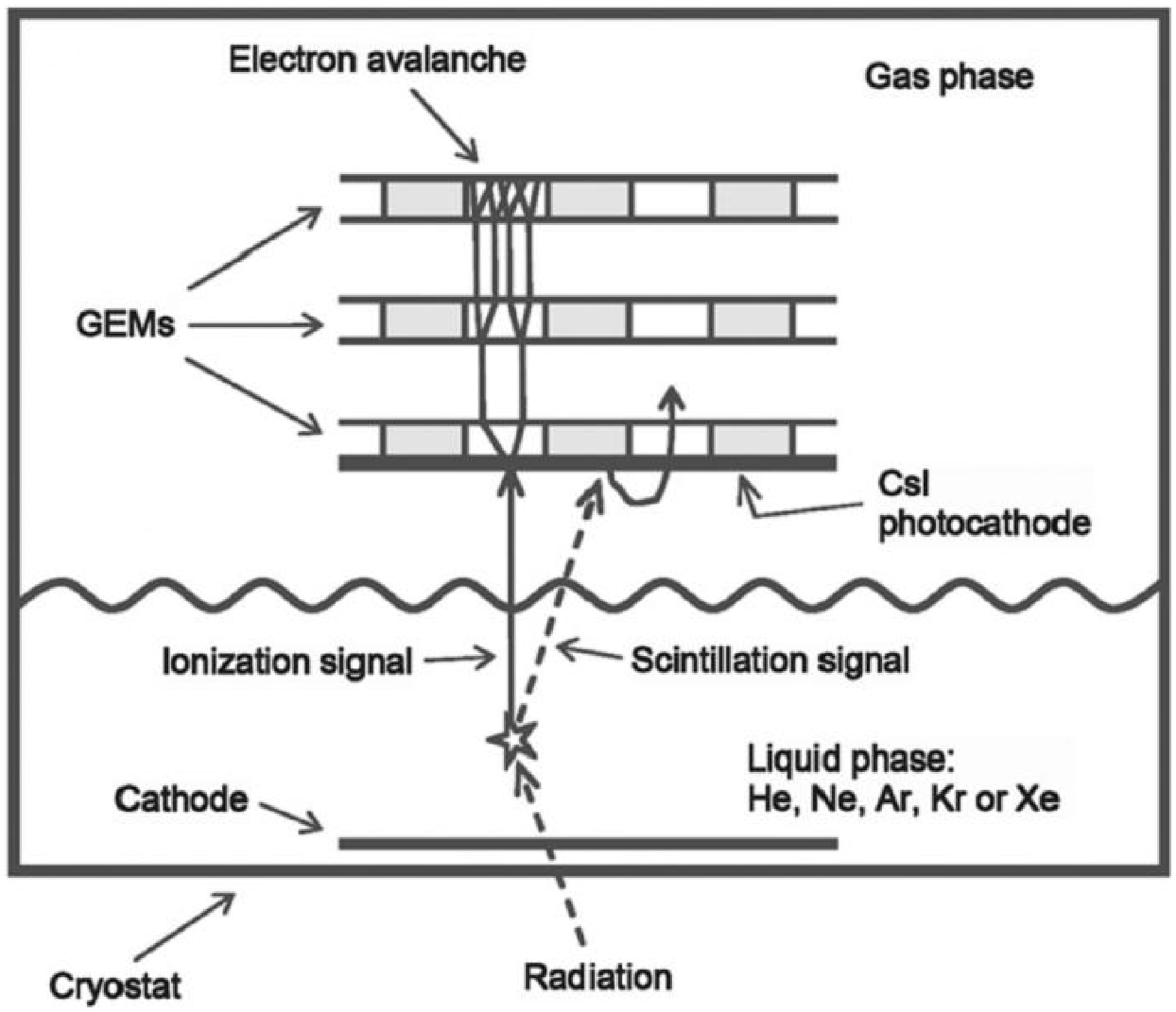}}
 \put(-3.0,37.0){ a) }
 \put(75.0,37.0){ b) }
 \end{picture}
\caption{ a) Operation principle of triple-GEM with $CsI$-PC in the Hadron Blind
Detector; 
b) Operation principle of the two-phase avalanche detector with a 
triple-GEM and reflective $CsI$-PC; both ionization and scintillation 
signals from the liquid are detected. }
\label{HadronBlind_Cryogenic}
\end{figure}

 Recent progress in the operation of cascaded GEM, Micromegas, THGEM and RETGEM GPMs at 
cryogenic temperatures (down to 80~K)~\cite{periale_ieeetns_52_927} 
and in two-phase mode~\cite{periale_nima_573_302,bondar_nima_581_241,bondar_nima_598_121} 
could pave the road towards their potential applications 
for the next generation neutrino 
physics and proton decay experiments~\cite{hagmann_ieeetns_51_2151,rubbia_0402110},
direct dark matter searches~\cite{rubbia_0510320},
Positron Emission Tomography (PET)~\cite{pet}, and
for noble-liquid Compton telescope, combined with a micro-PET 
camera~\cite{grignon_nima_571_142,2009_jinst4_p12008}.
The operation principle of the cryogenic two-phase avalanche detector with GEM readout 
is demonstarted in Fig.~\ref{HadronBlind_Cryogenic}b: 
the ionization produced in the noble liquid by radiation 
is extracted from the liquid into the gas phase by an electric field. 
 The multi-GEM detector, operated at cryogenic 
temperature in saturated vapor above the liquid phase,
can detect both the ionization signal, extracted from the liquid, and the 
scintillation signal, generated in the noble liquid by a particle~\cite{bondar_nima_581_241}.
The latter is achieved by depositing an UV-sensitive photocathode, 
namely CsI, on top of the first GEM (see Fig.~\ref{HadronBlind_Cryogenic}b). 
 The detection of both scintillation and ionization signals could allows
efficient background rejection in rare-event experiments;
in PET applications, the detection of scintillation signals could provide 
a fast trigger for coincidences between two collinear
gamma-quanta.

 $X$-ray counting and imaging detectors, based on MPGDs, are being used
for diffraction experiments at synchrotron radiation facilities~\cite{smith}
and could serve as a powerful diagnostic tool for magnetic fusion plasmas~\cite{jinst1_p09001_2006}.
An innovative GEM-based system, which combines fast imaging of $X$-ray emissions
with spectral resolution in the VUV range (0.2-10~$keV$), has been 
developed to study 2D dynamics of plasma
instabilities and to control core plasma position, both being crucial
issues for fusion researches~\cite{jinst1_p09001_2006}.
Another recent development is the construction of the spherical GEM detector for
parallax-free $X$-ray diffraction measurements, produced by the CERN/RD51 group
in collaboration with industry~\cite{pinto_2009IEEE}.
 In the rare-event experiments, MPGDs are used in searching for solar axions 
(CAST), where the expected signal comes from solar axions conversions 
into low-energy photons of 1-10~$keV$ range.
 Micromegas with high granularity anode elements can largely reduce the background 
event rate down to 5$\times10^{-5}$keV$^{-1}$cm$^{-2}$s$^{-1}$, 
exploiting its stability, good energy and spatial resolution~\cite{cast_nima604_15}.

 There are many applications of the Micromegas in
the neutron detection domain, which include neutron beam diagnostics~\cite{nima524_102},
inertial fusion experiments~\cite{nima557_648}, thermal neutron tomography~\cite{andriamonje}
and a sealed Picollo-Micromegas detector,
designed to provide in-core measurements of the neutron flux and
energy (from thermal to several MeV) in the nuclear reactor~\cite{picollo}.
 Neutrons can be converted into charged particles to detect ionization in Micromegas
by two means: either using the detector gas filling or
target with appropriate deposition on its entrance window.

 There were attempts to use GEM detectors for medical physics and 
portal imaging. In particular, GEM-based prototype was used to detect 
simultaneously the position of the therapeutic radiation beam ($\gamma$'s) 
and the position of the patient tumor, using $X$-rays, in order to provide feedback to the cancer 
treatment machine and to correct on-line position of the beam with respect to the 
patient~\cite{iacobaeus_ieeetns_48_1496,ostling_ieeetns_50_809}. 
In this device, the $X$-ray image was obtained using conversions in gas and 
the $\gamma$-beam profile was determined using solid converters 
placed in between several GEM foils.
Another medical application is the radiation therapy, 
which demands new on-line beam monitoring systems
with $\sim$~1~$mm$ spatial resolution and
3D dosimetry of delivered doses with an accuracy of $5~\%$.
 A scintillation light produced in the double-GEM detector during avalanche development
was detected by commercial CCD camera.
 With a 360~MeV $\alpha$-beam the scintillating GEM light signal at the Bragg peak
was only 4~$\%$ smaller than of the reference ionization chamber.
Consequently, GEM detector with CCD readout 
could become a feasible substitute for the Lanex scintillating screen, especially
for $\alpha$- or carbon-ion beams.

 Finally, MPGDs can be used for a variety of security related applications:
detection of dangerous cargo~\cite{arXiv_0911_3203}, radon detection in the air as 
an early warning of earthquakes~\cite{arXiv_1002_4732}
and UV sensitive early forest fire detection system~\cite{charpak_ieeetns_50_809,2009_jinst4_p12007}.
 Muon tomography, based on the measurement of multiple scattering of
cosmic ray muons traversing cargo of vehicles that contain high-Z materials, is a 
promising passive interrogation technique to detect hidden nuclear materials.
 Large area GEM detectors for precise tracking 
of atmospheric muons in combination with affordable electronics and readout system 
is being developed by one of the RD51 collaborating institutes~\cite{arXiv_0911_3203}.
 Among the planetary disasters the most common and often happening 
are the earthquakes and forest fires.
A sharp increase in the $Rn$ concentration before earthquakes has been observed
in the air regions associate to rocks and caves and its detection can serve as a basis
of an early earthquake warning system.
 The RETGEM-based device, capable to operate in the air and to 
continuously monitor the Rn concentration, is currently under development.
A network of such detectors can installed in the sensitive regions 
to provide daily assessment of the $Rn$ concentration~\cite{arXiv_1002_4732}. 
The same RD51 group is also developing a RETGEM-based
detector, which could be used for detection of forest fires 
at distances up to 1~$km$, compared
with a range of 200~$m$ for commercially available UV-flame 
devices~\cite{charpak_ieeetns_50_809,2009_jinst4_p12007}.

\section{Development of large area MPGDs}

 A big step in the direction of the industrial manufacturing of 
large-size MPGD's with size of few $m^2$ and spatial resolution typical of
silicon micro-strip devices ($30-50~\mu m$)
is the development of the new fabrication technologies -
single mask GEM~\cite{single_mask_GEM} and ``Bulk'' Micromegas~\cite{giomataris_nima_560_405}.
 Methods of their effective production and characterization are under investigation
within the RD51 collaboration.

 Recent developments on large area GEMs are focused on two new techniques 
to overcome the existing limitations: a single-mask technology and
a splicing method for GEM foils~\cite{single_mask_GEM}.
 The standard technique for creating the GEM hole pattern, involving accurate alignment of
two masks, is replaced by a single-mask technology to pattern only the top copper layer.
The bottom copper layer is etched after the polyamide, using the holes in the polyamide
as a mask.
 A single mask technique overcomes the cumburstone practice of alignment of
two masks between top and bottom films within 5-10~$\mu m$, 
which limits the achievable lateral size to $\sim 50~cm$.
In a splicing procedure, foils are glued over a narrow seam, obtaining a larger foil.
Both techniques were successfully implemented in the large-area
prototype of $66 \times 66~cm^2$ size, produced in collaboration with CERN TS-DEM workshop, 
shown in Fig.~\ref{Large_area_MPGD}a. 
 Single mask GEM seems to be much more suitable for industrial
large scale production than standard GEM technology.

 The basic idea of the ``Bulk'' Micromegas technology is to built the whole detector in a single process:
the anode plane with copper strips, a photo-imageable polyamid film
and the woven mesh are laminated together at a high temperature forming a single 
object~\cite{giomataris_nima_560_405}.
 At the end, the micromesh is sandwiched between two layers of insulating material,
which is removed after UV exposure and chemical development.
 This ``Bulk'' Micromegas technique has been recently extended
to produce large area prototypes in CERN TS-DEM workshop, up to $150 \times 50~cm^2$
in a single piece (see Fig.~\ref{Large_area_MPGD}b)~\cite{alexopouloc_jinst_atlas}.

 The THGEM and RETGEM technologies could offer economically interesting solution for a
single-photon Cherenkov imaging counters: 
namely good spatial and time resolutions, 
large gains (single photo-electron sensitivity), relatively low mass
and easy construction - thanks to the intrinsic robustness of the PCB electrodes.
 To advance future developments of GPMs, one of the key ingredients
is the possibility to produce, with good quality and yield, 
THGEM-like foils of large surfaces coated with $CsI$ UV-sensitive photocathodes.
 There is an ongoing $R\&D$ within the RD51 collaboration to optimize
THGEM geometry and to study charging up effects, gain stability 
and maximum achievable gain (similar to GEMs).
Fig.~\ref{Large_area_MPGD}c shows a first $60 \times 60~cm^2$ THGEM foil
produced in cooperation with industry (ELTOS, Italy)~\cite{tessarotto}.

\setlength{\unitlength}{1mm}
\begin{figure}[bth]
 \begin{picture}(55,55)
 \put(2.0,-5.0){\includegraphics{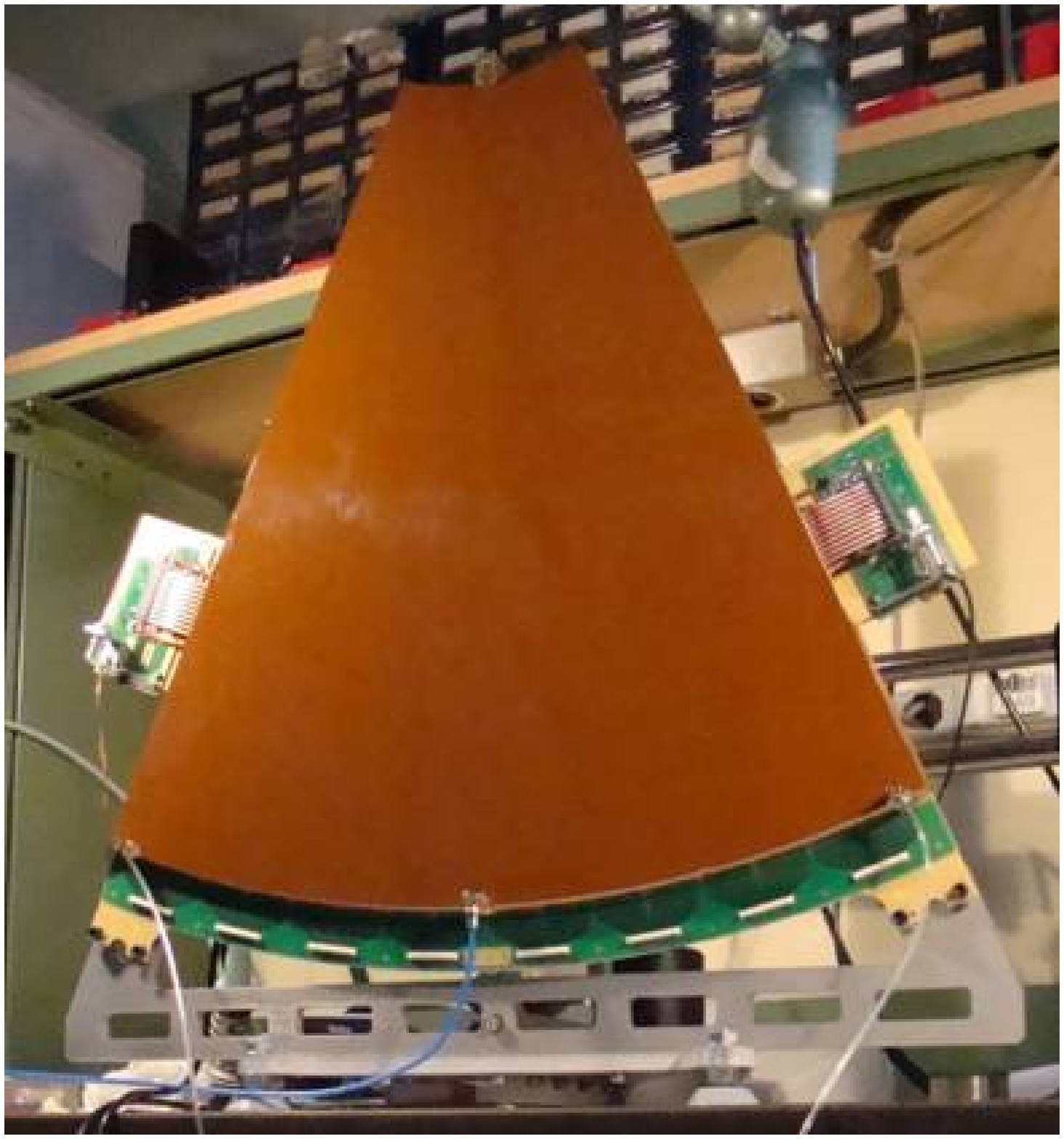}}
 \put(54.0,-5.0){\includegraphics{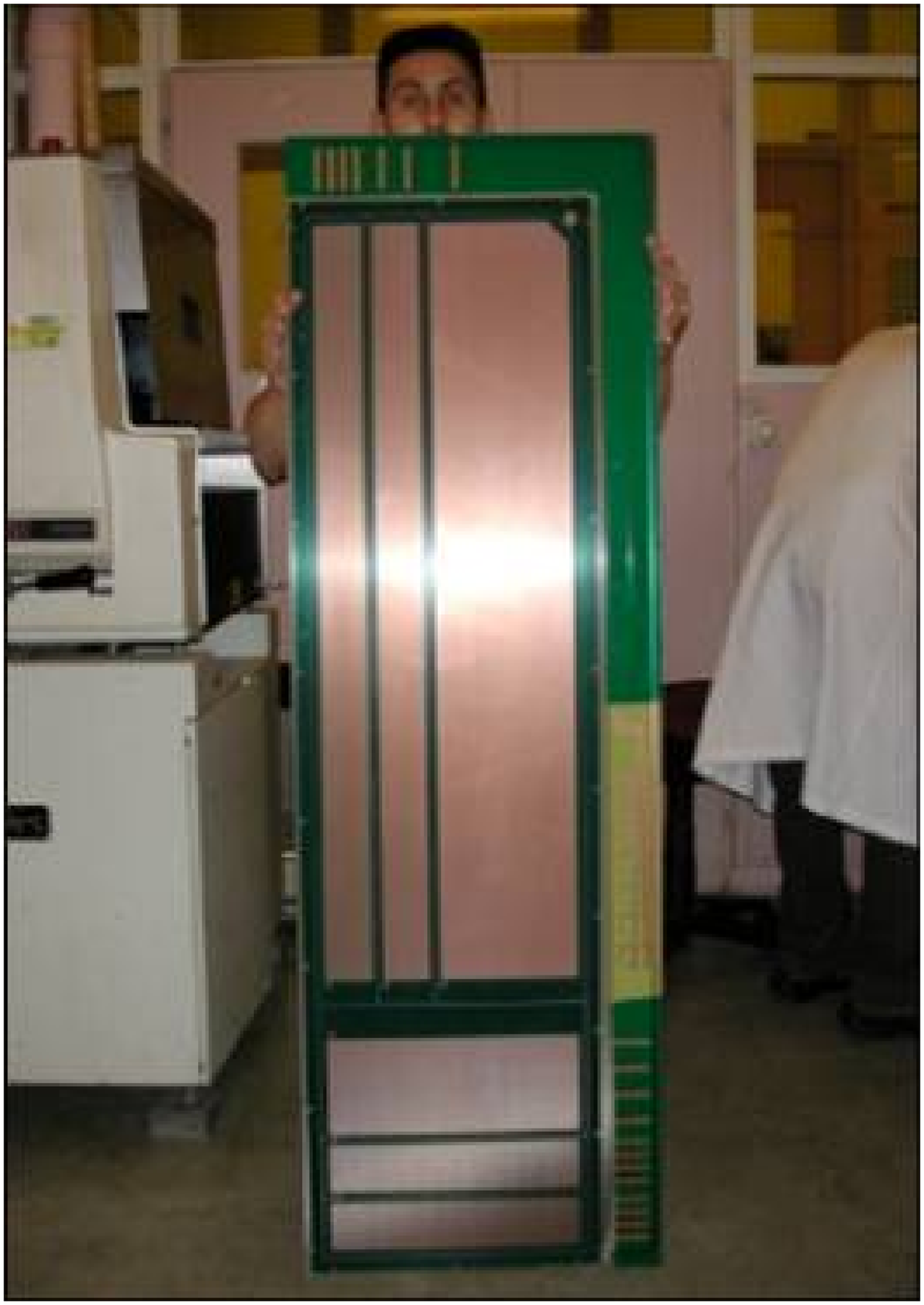}}
 \put(105.0,-5.0){\includegraphics{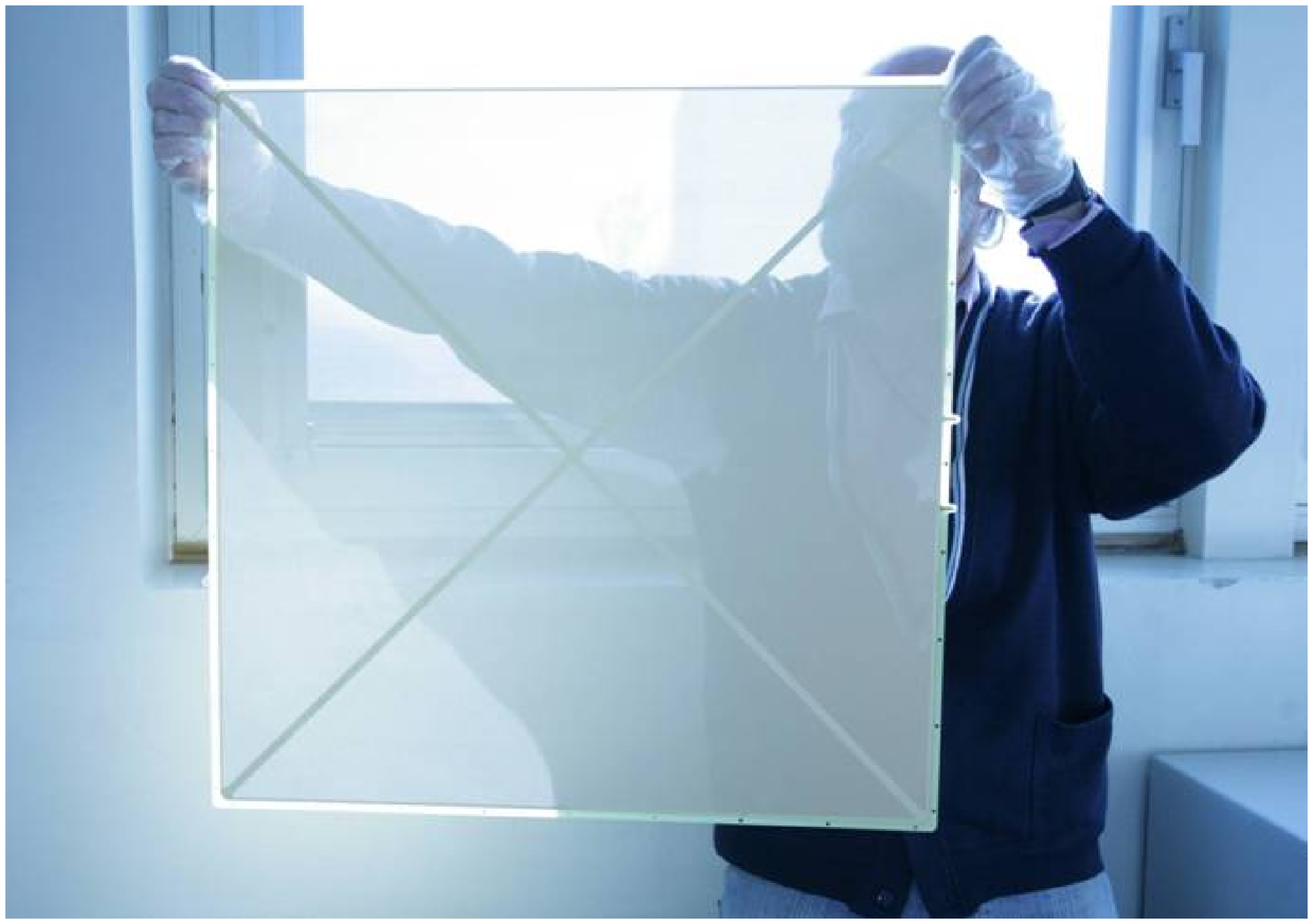}}
 \put(-1.0,40.0){ a) }
 \put(50.0,40.0){ b) }
 \put(102.0,40.0){ c) }
 \end{picture}
\caption{ a) The triple-GEM prototype of $66 \times 66~cm^2$ active area,
using single-mask GEM technology, for TOTEM experiment;
b) Large area ``Bulk'' Micromegas prototype of $40 \times 130~cm^2$ size
for the sLHC ATLAS muon system upgrade;
c) First $60 \times 60~cm^2$ THGEM foil produced for the COMPASS RICH upgrade.}
\label{Large_area_MPGD}
\end{figure}

 The development of large-area position-sensitive photon detectors, with visible sensitive photocathodes
(e.g. bialkali), could lead to numerous spin-off, beyond the Cherenkov light imagers.
Most commonly-used in the visible range are vacuum photomultipliers (PMTs), with rather limited 
module size and bulky geometry due to mechanical constraints 
on the glass vacuum envelope. 
 A possible alternative is to use gas-filled photomultipliers at atmospheric pressure;
a proof of principle of visible-sensitive GPM was demonstrated
with MPGDs~\cite{chechik_nima595_116,lyashenko_jinst4_p07005}.
Future studies with large-area detection and visible-sensitive PC
are being planned within the RD51 collaboration.

\section{Pixel readout for micro-pattern gas detectors}

 Coupling of the micro-electronics industry together with advanced PCB technology 
has been very important for the development of modern gas detectors with increasingly 
smaller pitch size. 
The fine granularity and high-rate capability of micro-pattern devices can be further exploited 
by introducing high-density pixel readout with a size corresponding to the intrinsic 
width of the detected avalanche charge.
 However, for a pixel pitch of the order of 100~$\mu m$, technological constraints 
severely limit the maximum number of channels that can be brought to the external 
front-end electronics. 
 While the standard approach to readout MPGDs is a segmented strip or pad plane with front-end 
electronics attached through connectors from the backside, an attractive possibility is to 
use CMOS pixel chip, assembled directly below the GEM or Micromegas amplification 
structures, and serving  as integrated device hosting the input pad, the preamplification 
and the digitization of the 
signals~\cite{nature411_2001,bellazzini_NIMA535_477,campbell_NIMA540_295,bamberger_NIMA573_361}.
Due to the strong electric field in the GEM or Micromegas
amplification structures, primary electron initiates avalanche, and the resulting
charge signal activates the preamp-shaper-discriminator circuitry,
integrated in the underlying active layers of the CMOS technology.
 Using this approach, gas detectors can reach the level of integration,
compactness and resolving power typical of 
solid-state pixel devices.

\setlength{\unitlength}{1mm}
\begin{figure}[bth]
 \begin{picture}(40,40)
 \put(13.0,-5.0){\includegraphics{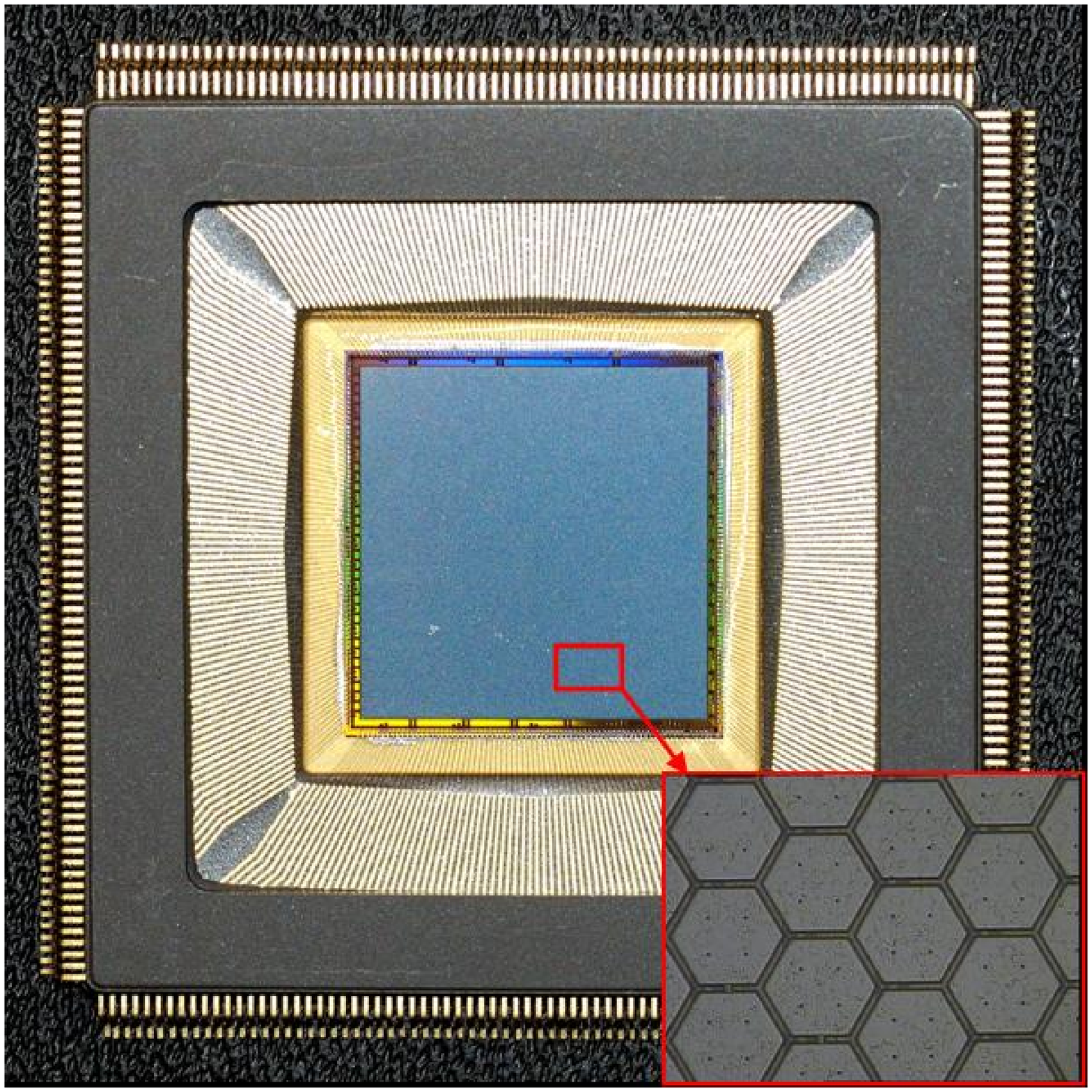}}
 \put(80.0,-5.0){\includegraphics{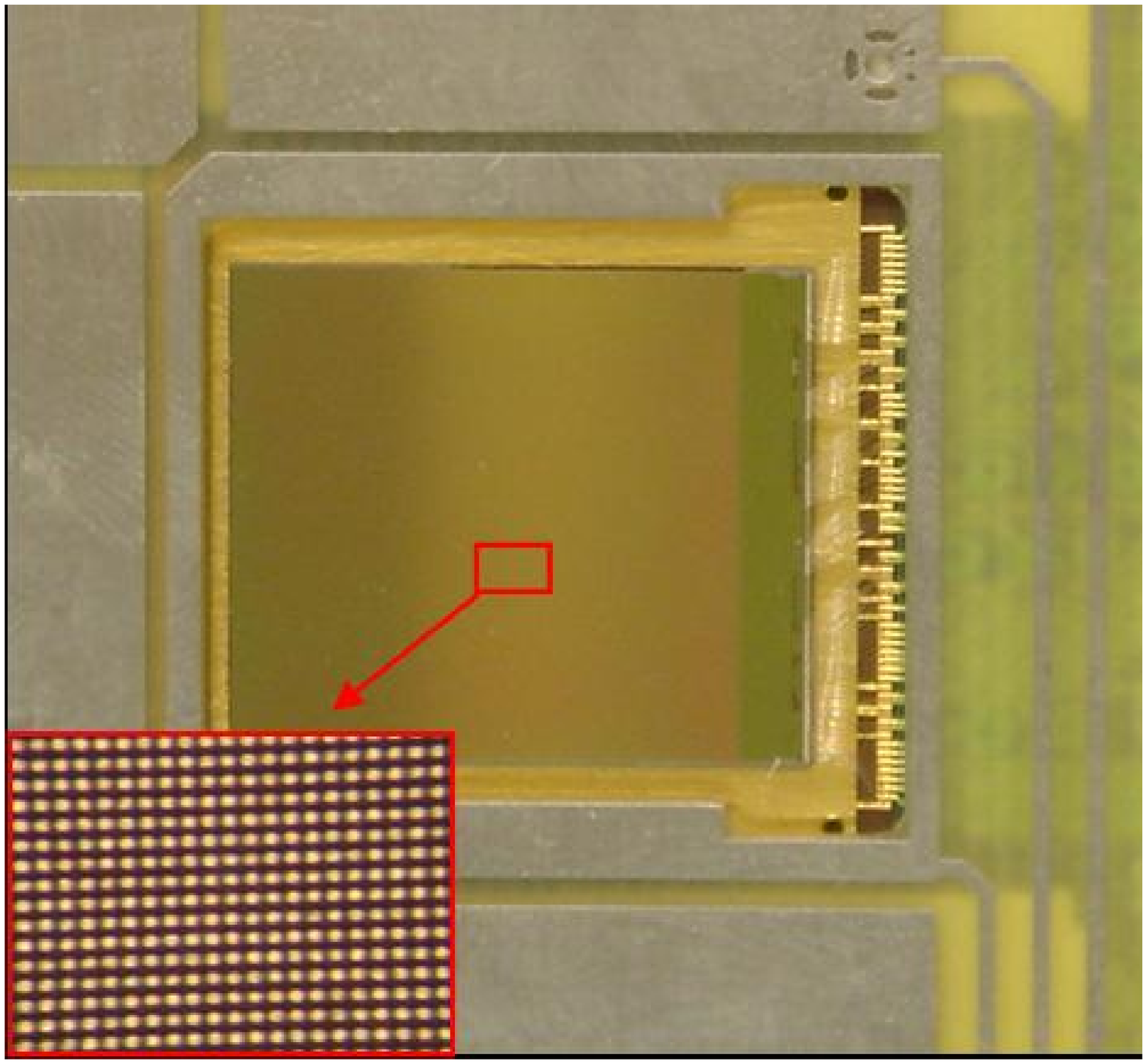}}
 \put(9.0,35.0){ a) }
 \put(76.0,35.0){ b) }
 \end{picture}
\caption{ a) Photo of the analog CMOS ASIC with hexagonal pixels,
bonded to the ceramic package;
b) Photo of the Medipix2 chip; a 25~$\mu m$ wide conductive bump bond openings,
used for electron collection, are seen as a matrix of dots.}
\label{CMOSASIC_photo}
\end{figure}

 The combination of GEM detector and an analog, low-noise and high 
granularity (50~$\mu m$ pitch) CMOS pixel ASIC,
comprising pixellated charge collecting anode and readout electronics,
can bring large improvement in sensitivity, at least 2 orders of magnitude,
compared to traditional $X$-ray polarimeters (based on Bragg diffraction or
Compton scattering)~\cite{spie_4843_372,spie_4843_394,muleri_nima584_149}.
 Moreover, scattering polarimeters are practically insensitive below 5~keV
and are background limited, while Bragg crystal polarimeters are efficient
only around a narrow band fulfilling the Bragg condition.
In contrary, the pixel readout of MPGDs
allows to reconstruct individual
low-energy (2-3~keV) photo-electron tracks with a length as short as
few hundred microns; the total charge collected in the pixels
is proportional to the $X$-ray energy.
 The high detector granularity allows to recognize the initial part of the track,
before it is distorted by Coulomb scattering,
through the localization of absorption point of the photon and then
to estimate the photo-electron emission direction.
 Finally, the degree of $X$-ray polarization is computed
from the distribution of reconstructed track angles, since the
photo-electron is emitted mainly in the direction of photon electric field.
 Three ASIC generations of increased complexity and size, reduced pitch and
improved functionality have been designed and 
built~\cite{bellazzini_NIMA535_477,bellazzini_NIMA560_425,bellazzini_NIMA566_552,bellazzini_NIMA572_160}
(see Fig.\ref{CMOSASIC_photo}a).
 The third ASIC version, realized in 0.18~$\mu m$ CMOS technology,
includes self-triggering capability and has
105600 hexagonal pixels with a 50~$\mu m$ pitch,
corresponding to an active area of 15$\times$15 $mm^2$.
 Single GEM detector coupled to such a CMOS pixel chip
can convert photons in the energy range from a few keV up to tens of keV,
by choosing the appropriate gas mixture, and is able to simultaneously 
produce high resolution images ($50~\mu m$), moderate spectroscopy 
(15$\%$ FWHM at 6~keV) and fast timing (30~ns) signals.
 At the focal plane of the large area mirror like XEUS telescope,
the novel device will allow to perform energy-resolved
polarimetry at the level of few $\%$ on many galactic and
extragalactic sources with photon fluxes down to one milliCrab 
in one day~\cite{bellazzini_NIMA566_552,bellazzini_NIMA572_160}.

 The original motivation of combining a MPGD with Medipix2~\cite{ieee_tns_49_2279} 
and Timepix~\cite{nima581_485} chips 
was the development of a new readout system for a large 
TPC at the future Linear Collider.
 The digital Medipix2 chip was originally designed for single photon counting by means of a
semiconductor $X$-ray sensor coupled to the chip. 
 In gas detector applications, the chip is placed in the gas volume 
without any semiconductor sensor, with GEM or Micromegas amplification
structure above it~\cite{campbell_NIMA540_295,colas_nima535_506,bamberger_NIMA573_361}. 
 Approximately 75~$\%$ of every pixel in the Medipix2 matrix is covered
with an insulating passivation layer.
Hence, the avalanche electrons are collected on the metalized input pads,
exposed to the gas (see Fig.~\ref{CMOSASIC_photo}b).
 The Timepix chip, which is a modification of the Medipix2 chip,
yields timing and charge measurements as well as precise
spatial information in 3D of individual electron clusters.
 The Timepix ASIC sensitive area is arranged as a square matrix of 256 x 256 pixels of 
55 x 55 $\mu m^2$ size, resulting in a total detection area of $\sim 14\times 14~mm^2$,
which represents 87~$\%$ of the entire surface area. 
The chip is designed and manufactured in a six-metal 0.25~$\mu m$ CMOS technology and 
contains $\sim$50 million transistors.
 Each pixel in the chip matrix can be programmed to record either the arrival time 
of the avalanche charge signal with respect to an external shutter (``TIME'' mode) or
the 14-bit counter is incremented as long as the signal remains above the threshold
(Time Over Threshold ``TOT'' mode), thus providing pulse-height information.
 The operation of MPGD with Timepix chip has demonstrated the possibility
to reconstruct 3D-space points of individual primary electron clusters with $\sim30-50~\mu m$
spatial resolution and event-time resolution with $ns$ 
precision (see Fig.~\ref{timepix_image}b)~\cite{bamberger_nima581_274,arXiv_07092837,jinst_11015_2009}. 
 Thanks to these developments, a micro-pattern device with
CMOS readout can serve as a high-precision 
``electronic bubble chambers''~(see Fig.~\ref{timepix_image}a).

\setlength{\unitlength}{1mm}
\begin{figure}[bth]
 \begin{picture}(45,45)
 \put(0.0,-5.0){\includegraphics{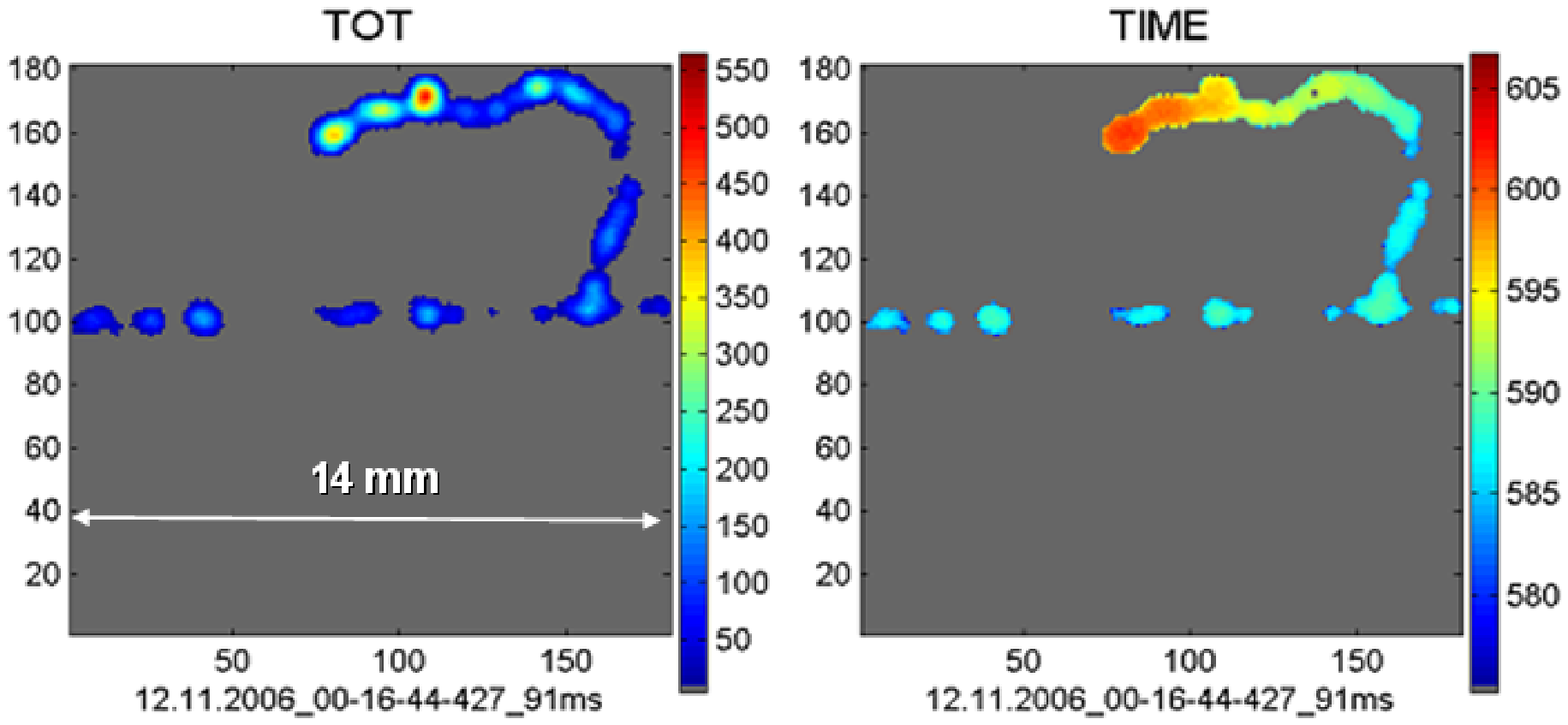}}
 \put(105.0,-5.0){\includegraphics{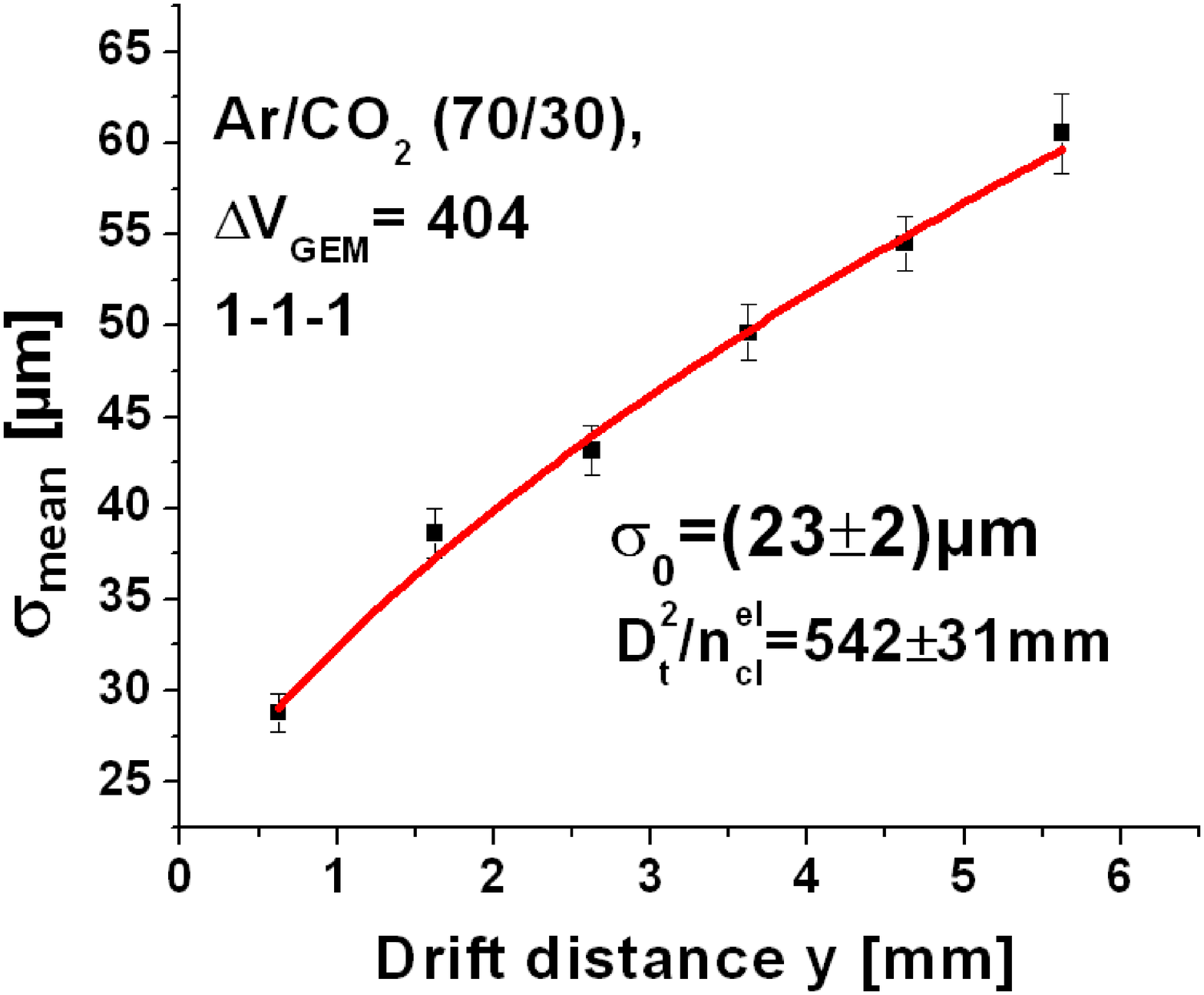}}
\put(-2.0,40.0){ a) }
\put(105.0,40.0){ b) }
 \end{picture}
\caption{ a) Image of the 5~GeV electron track recorded at the DESY testbeam with
triple-GEM detector and TimePix CMOS ASIC operated in the mixed mode:
every second pixel is operated in the ``TOT''/''TIME'' modes in the ``chess board'' fashion.
The $x$, $y$-axis represent chip sensitive area, obtained by mapping the original data
(matrix of 256 $\times$ 256 pixels of 55~$\mu m$ pitch) into a 181 $\times$ 181 pixels
matrix with a pitch of 78~$\mu m$.
The color is the measure of the time-over-threshold and drift time information, respectively;
b) Spatial resolution as a function of drift distance
in $Ar/CO_2$(70:30) mixture and triple-GEM/Medipix2 detector.}
\label{timepix_image}
\end{figure}

 An elegant solution for the construction of the Micromegas with pixel
readout is the integration of the amplification grid and CMOS chip
by means of an advanced ``wafer post-processing'' technology~\cite{chefdeville_NIMA556_490}.
With this technology, the structure of thin (1 $\mu m$) aluminum grid
is fabricated on top of an array of insulating (SU-8) pillars of typically
50 $\mu m$ height, which stand above the CMOS chip, forming an integrated readout of the gaseous
detector (see Fig.~\ref{InGrid_SiProt}a). 
This technology is called ``InGrid'' 
(developed by the collaboration of MESA+ institute of the University of Twente and NIKHEF) and 
allows an accurate alignment 
of hexagonal grid holes ($\sim~60~\mu m$ pitch) with pixel input pads, 
sub-micron precision in the grid hole diameter and a constant 
thickness of the avalanche gap, which results in a uniform gas gain.
The process uses standard photo-lithography and wet etching techniques and is CMOS compatible. 
It can be used to equip both single chips and chip wafers with Micromegas grid.
Amongst the most critical items that may affect the long-term operation of ``InGrid'' concept 
is the appearance of breakdown across the 50-100~$\mu m$ amplification gap. 
This is a critical issue for Micromegas with CMOS readout, where a multiplication grid 
is placed directly above the naked Timepix chip, since for single stage 
amplification discharges in presence of heavy ionizing particles can not be 
completely eliminated. 

One way to achieve protection is to cover the CMOS chip with a thin layer (3 to 20 $\mu m$) of
highly resistive amorphous silicon or silicon nitride ($Si_3 N_4$)
deposited directly on top of the Timepix ASIC~\cite{2006IEEE_aarts,2007IEEE_bosma}.
This protective cover, called ``SiProt'', 
intends to protect the chip from high instantaneous spark currents, which could destroy the chip, 
and from the the evaporation of the Timepix metal input pads during sparks (see Fig.~\ref{InGrid_SiProt}b). 
 The performance of the ``InGrid'' detector with 4~$\mu m$ ``SiProt'' layer 
is illustrated in Fig.~\ref{InGrid_SiProt}c for 2~GeV electron beam at DESY.
Due to its high sensitivity, ``InGrid'' detector can resolve single primary 
electrons~\cite{campbell_NIMA540_295,carballo_nima583_42}. 
The color is a measure of the electron arrival time in ``TOT'' mode, red color
corresponds to the primary ionization electrons produced close to the  chip surface.
The spread of the clusters in Fig.~\ref{InGrid_SiProt}c indicates that the diffusion increases 
with the distance from the chip.
The time range in $He/iC_4 H_{10}$ mixture is larger than in $Ar/iC_4 H_{10}$,
which indicates lower drift velocity in $He$-based mixture, while the 
primary ionization density (number of active pixels) is higher in $Ar$-based mixture~\cite{2009IEEE_fransen}.
 The ``InGrid'' device with a narrow (1-1.5~$mm$) drift gap (``GOSSIP'') is currently proposed 
as one of the possible upgrade options 
for the ATLAS tracking system at sLHC, aiming for a spatial precision of about 
20~$\mu m$~\cite{rd51note_2009_006}.

\setlength{\unitlength}{1mm}
\begin{figure}[bth]
 \begin{picture}(40,40)
 \put(3.0,-4.0){\includegraphics{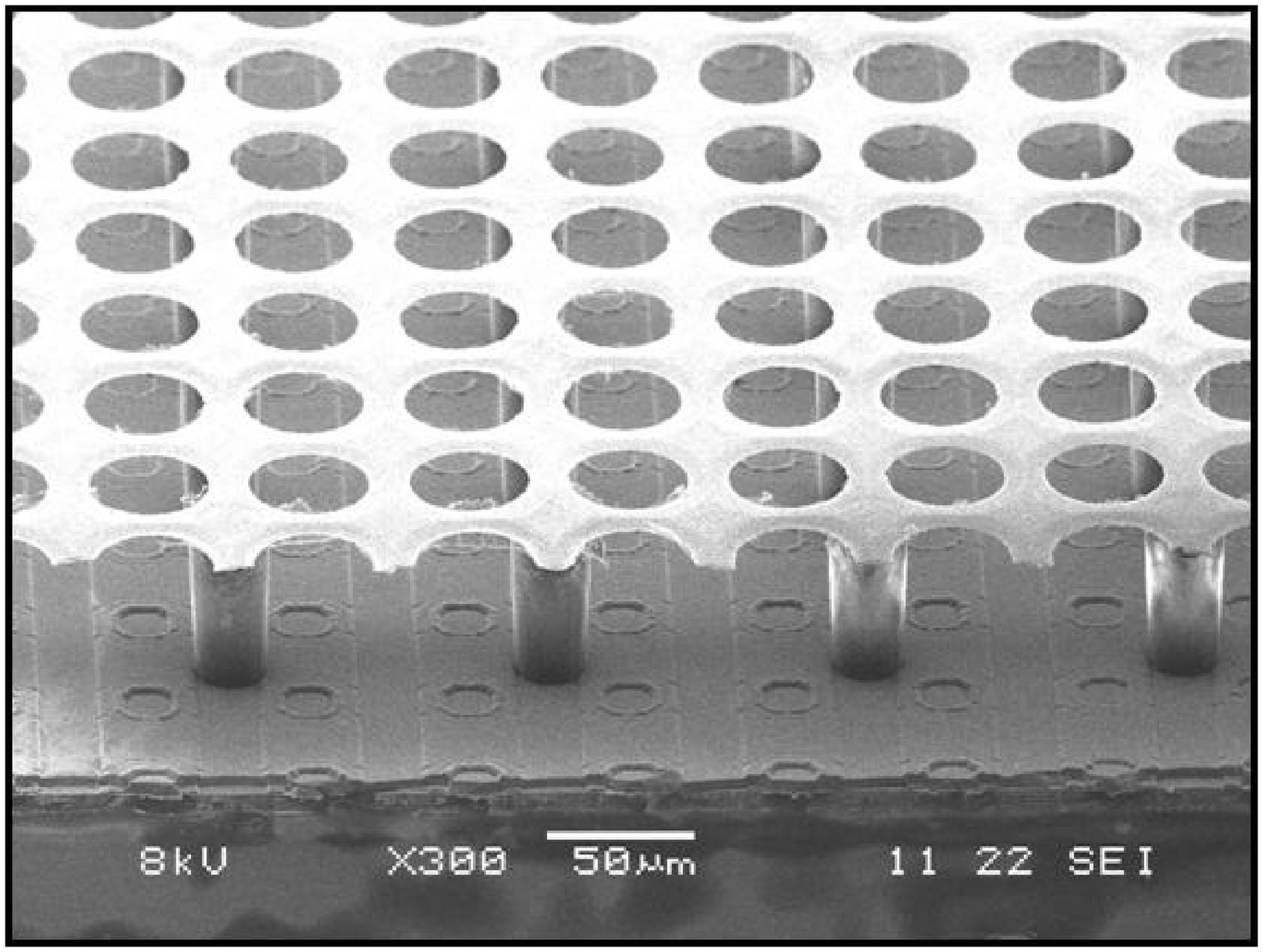}}
 \put(59.0,-4.0){\includegraphics{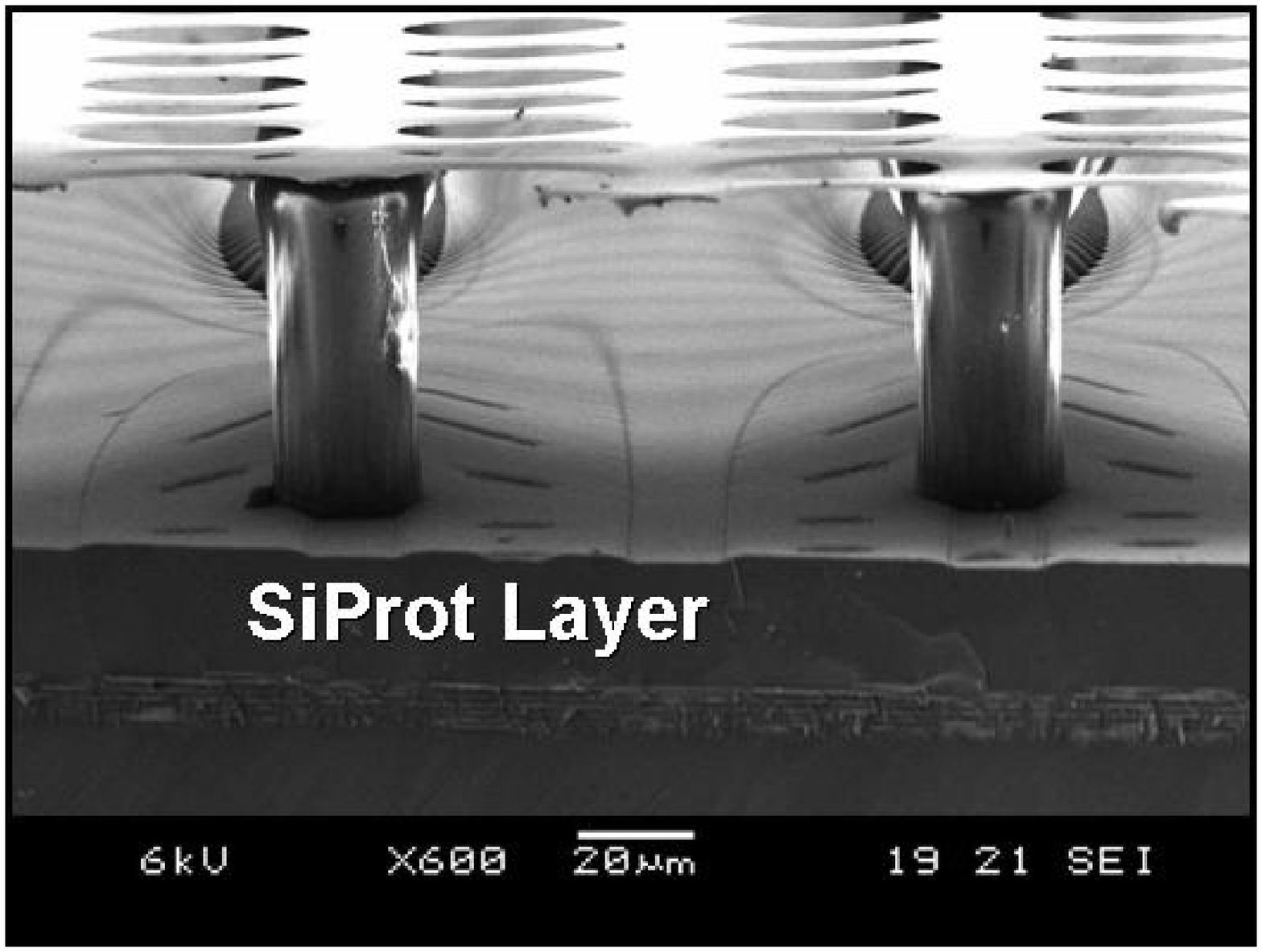}}
 \put(111.0,-4.0){\includegraphics{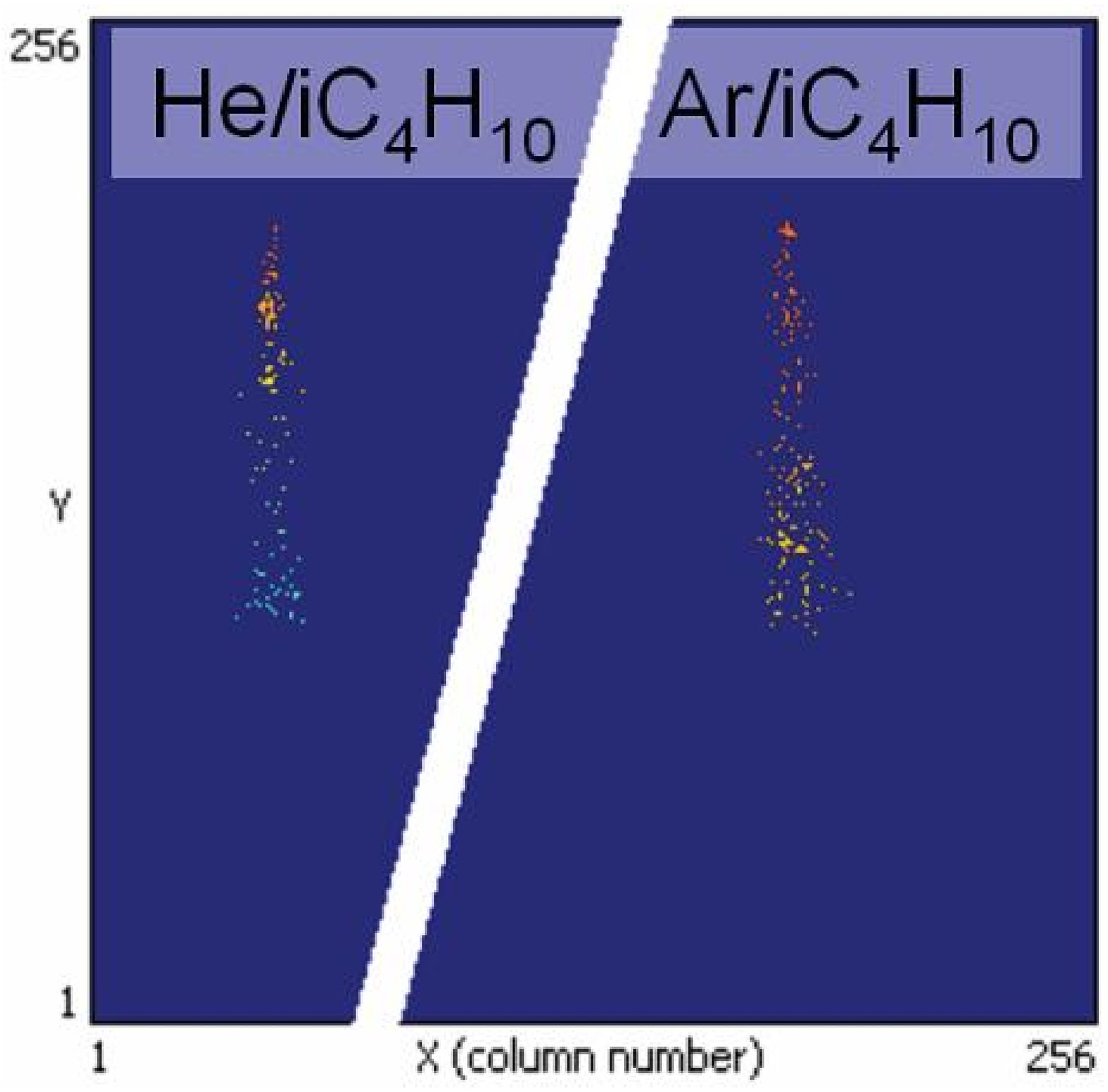}}
 \put(-1.0,37.0){ a) }
 \put(55.0,37.0){ b) }
 \put(109.0,37.0){ c) }
 \end{picture}
\caption{ a)  The Scanning Electron Microscopy (SEM) image of the ``InGrid''
detector: a Timepix chip, resistive silicon-nitride ($Si_3 N_4$) protection layer and
insulating SU-8 pillars supporting a perforated Al grid. 
The pillars are placed at the intersections of four adjacent pixels; 
the circular holes in the grid are centered onto the input pad of each pixel;
b) Photo of the ``InGrid'' structure with 7~$\mu m$ $Si_3 N_4$ protection layer;
c) Images of 2~GeV electron tracks recorded with ``InGrid'' detector in $Ar/C_4H_{10}$
and $He/C_4H_{10}$ mixtures in ``TIME'' mode;  the color is a measure of the arrival time 
of electrons.}
\label{InGrid_SiProt}
\end{figure}

  One of the most exciting future applications of GEM and Micromegas devices with
CMOS multi-pixel readout could be position sensitive single photon detection.
  Recently, a UV photo-detector based on a semitransparent CsI photocathode followed
by the fine-pitch GEM foil, that matches the pitch of a pixel ASIC (50~$\mu m$),
has shown excellent imaging capabilities~\cite{bellazzini_nima_581_246}.
 The photoelectron emitted from CsI layer drifts into a
single GEM hole and initiates an avalanche, which is then collected on the
pixel CMOS analog chip.
 Due to the high granularity and large $S/N$ of the read-out system,
the ``center of gravity'' of the single electron avalanche corresponds to
the center of GEM hole.
 Accumulating thousands of such events produces
``self-portrait'' of the GEM amplification structure,
shown in Fig.~\ref{Ingrid_UV_CsI}~(a) and
allows to achieve superior single electron avalanche 
reconstruction accuracy of 4~$\mu m$~$rms$~\cite{bellazzini_nima_581_246}.
 Another monolithic gaseous UV-photon imaging device is based on the ``InGrid''
concept and reflective CsI photocathode deposited on Micromegas grid.
The photocathode deposition and its operation is found to be adequate
in combination with both the ``InGrid'' detector and Timepix chip.
Fig.~\ref{Ingrid_UV_CsI}b shows image recorded with a steel mask placed in front
of the ``InGrid'' detector under UV-photon irradiation.
 The detector operated reliably in $He/iC_4 H_{10}$ mixture with
single electron extraction efficiency of $\sim 50~\%$ and a 
maximum achievable gain of $\sim 6\times 10^4$.
 These results encourage further developments towards
high-resolution UV photon devices, based on combination of MPGD with CMOS pixel readout.
Futhermore, being made of UHV-compatible materials, ``InGrid''
concept could be used in cascaded visible light-sensitive imaging gas photomultipliers.

\setlength{\unitlength}{1mm}
\begin{figure}[bth]
 \begin{picture}(40,40)
 \put(20.0,-3.0){\includegraphics{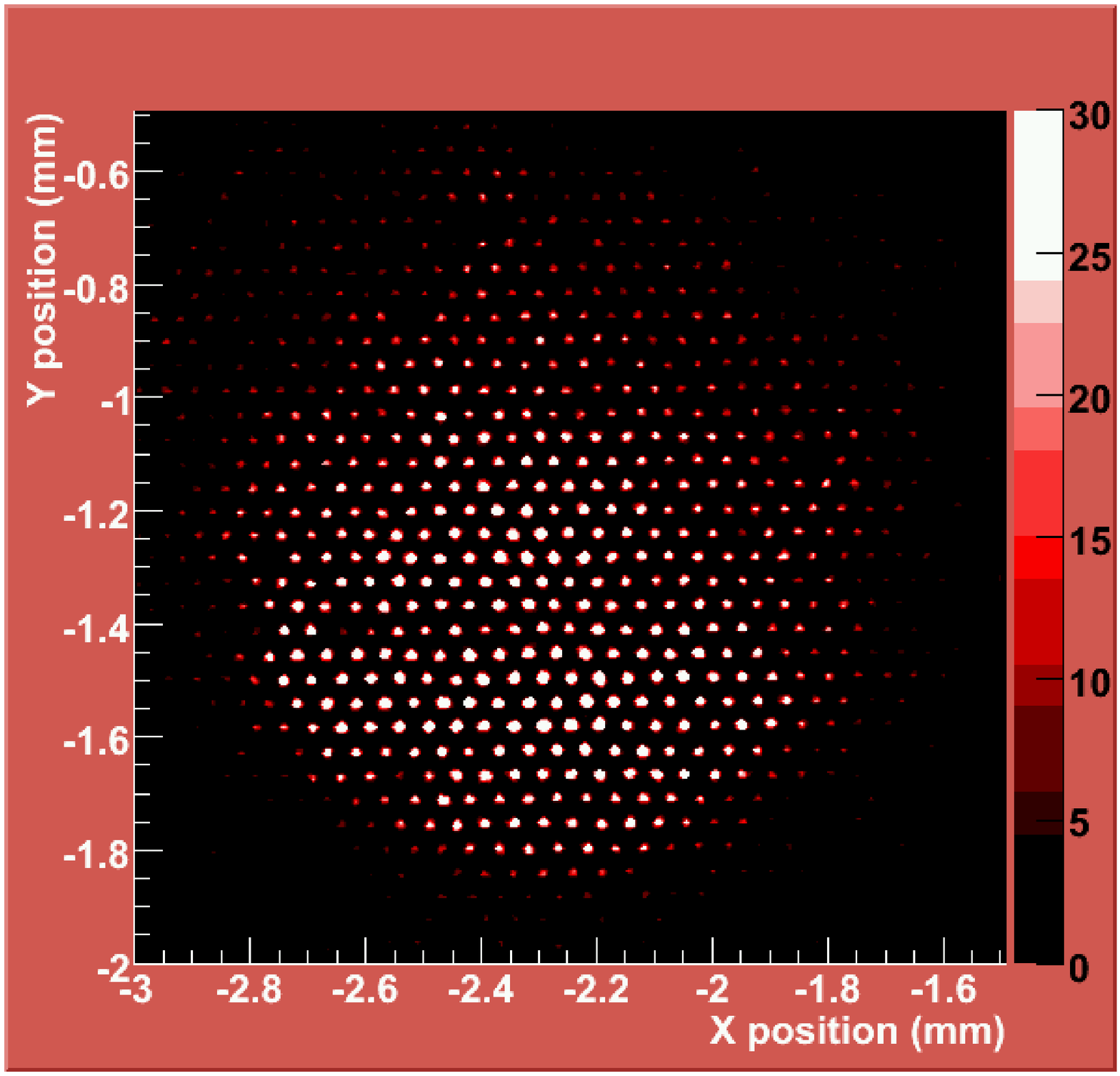}}
 \put(87.0,-7.0){\includegraphics{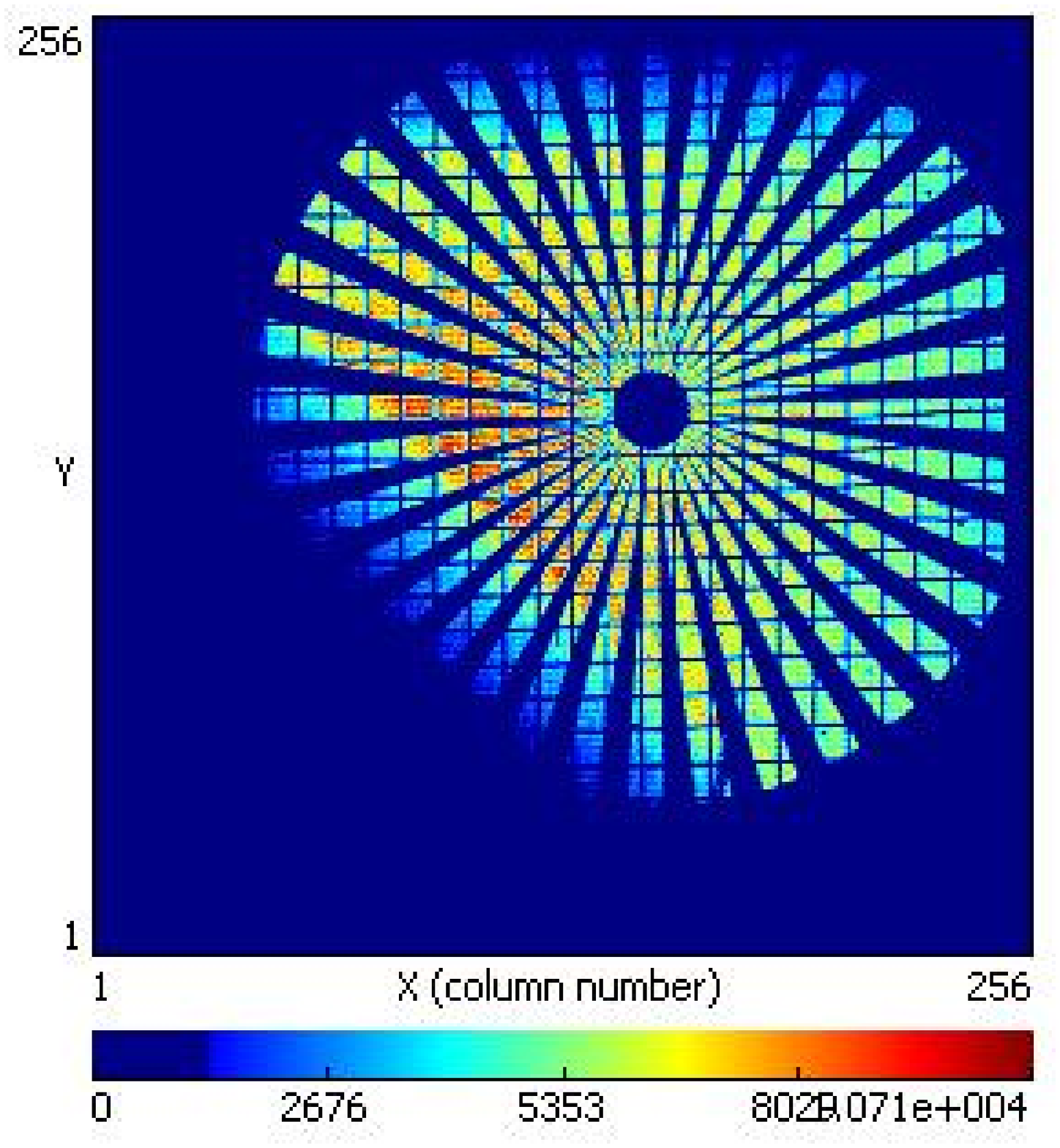}}
 \put(14.0,35.0){ a) }
 \put(80.0,35.0){ b) }
 \end{picture}
\caption{ a) Cumulative map of hundreds of thousands events from the UV light
conversion in the semitransparent CsI photocathode of the detector
producing a kind of ``self-portrait'' of the GEM amplification structure.
The white spots correspond to the individual GEM holes at 50~$\mu m$ pitch;
b) 2D image of the steel mask under irradiation of the ``InGrid'' detector
with reflective CsI photocathode in $He/iC_4 H_{10}$~(80:20) mixture.}
\label{Ingrid_UV_CsI}
\end{figure}

  A key point that has to be solved to allow CMOS pixel readout of MPGD for applications 
in various fields of fundamental science and beyond is the production of large area detectors.
Presently prototypes under construction rely at maximum on $2\times 4$ chips integrated 
readout~\cite{2009IEEE_lentdecker,kaminski_rd51_november2009}.
  Going to large surface detector requires solution for the dead space on one
side of the Timepix chip, where the electrical connections enter the 
chip (see Fig.~\ref{CMOSASIC_photo}b). 
A first solution relies on etching of ``through-silicon vias'' on the Timepix chip 
that allows to bring out the signals and services on the backside of the chip
using ``through-wafer-vias'' technology~\cite{takahashi,heijne}. 
This is known as ``via-last'' operation: 
the existing Timepix wafers are  modified by etching ``in-vias'' connections 
after the CMOS wafer production step is completed.
In order to move from 3-side tileable detectors to 4-side, all common circuitry 
(bias, converters, ...) needs to be spread over the chip area which is nowadays
become possible thanks to the progress in 3D micro-electronics, where the different functions 
can be realized on different tiers and connected thanks to ``through-silicon vias'' 
and inter-connects.
However, a major R$\&$D effort is 
required in the future to fully exploit this potential.

 Properly integrated into large systems (including development of large area 
MPGD based on the integrated multi-chip CMOS ASICs),
the pixel readout may open new opportunities for
an advanced Compton Telescope~\cite{takada_nima546_258,miuchi_nima576_43}, 
detection of Axions and Weakly Interacting Massive Particles~\cite{ahlen_ijmp25_2010_1}, 
neutrino-less double beta decay experiments~\cite{carballo_jinst5_p02002}
and 3D imaging of nuclear recoils.
 The primary advantage of a gaseous tracker with pixel readout is that the direction of the
Compton electron or the low energy nuclear recoil can be reconstructed
far more accurately than in any other detection medium.

\section{Summary}

Radiation detection and imaging with gas-avalanche detectors, capable of economically 
covering large detection volumes with a low material budget, 
have been playing an important role in many fields. 
While extensively employed at the LHC, RHIC, and other advanced HEP experiments, 
present gaseous detectors (wire-chambers, drift-tubes, resistive-plate chambers 
and others) have limitations which may prevent their use in future experiments. 

The possibility of producing micro-structured semi-conductor devices 
(with structure sizes of tens of microns) and corresponding highly 
integrated readout electronics led to the success of semi-conductor 
(in particular silicon) detectors to achieve unprecedented space-point 
resolution. 
 Micro-pattern  gas-amplification structures now open the 
possibility to apply the same technology to gaseous detectors and enable 
a plethora of new detector concepts in science and industry,
in medical and commercial applications.
Microelectronics needs to keep up with the state-of-the-art developed by the microelectronics 
industry and the interconnection between electronics and detectors needs to be improved to 
reduce the complexity and the material of interconnect technologies.

The RD51 collaboration at CERN, approved in December 2008, aims at facilitating the development 
of advanced gas-avalanche detector technologies and associated electronic-readout systems, 
for applications in basic and applied research. 
The main objective of the R$\&$D program is to advance technological development 
and application of micro-pattern gaseous  detectors.
The RD51 collaboration involves $\sim$~430 authors, 73 Universities and Research Laboratories 
from 25 countries in Europe, America, Asia and Africa. 
All partners are already actively pursuing either basic- 
or application-oriented R$\&$D involving a variety of MPGD concepts.
The collaboration established common goals, like experimental and simulation tools, 
characterization concepts and methods, common infrastructures at test beams and 
irradiation facilities, and infrastructure for MPGD production.
More information can be found at the RD51 collaboration webpage~\cite{rd51webpage}.

\end{document}